\documentclass[12pt,titlepage]{article}

\usepackage{amssymb,amsfonts,amsmath,amsthm,mathtools,bm}
\usepackage{bbm}

\usepackage[dvipsnames]{xcolor}
\let\SetColor\color
\usepackage{etoolbox}
\makeatletter
\AtBeginEnvironment{align*}{\let\SetColor\@gobble \color{black}}
\AtBeginEnvironment{align}{\let\SetColor\@gobble \color{black}}
\AtBeginEnvironment{aligned*}{\let\SetColor\@gobble \color{black}}
\AtBeginEnvironment{aligned}{\let\SetColor\@gobble \color{black}}
\AtBeginEnvironment{equation*}{\let\SetColor\@gobble \color{black}}
\AtBeginEnvironment{equation}{\let\SetColor\@gobble \color{black}}
\AtBeginEnvironment{underline}{\let\SetColor\@gobble \color{black}}
\AtBeginEnvironment{tabular}{\let\SetColor\@gobble \color{black}}
\makeatother

\definecolor{MyBlue}{RGB}{0,91,148}
\definecolor{MyRed}{RGB}{200,15,62}
\definecolor{MyPurple}{rgb}{0.44, 0.16, 0.39}
\definecolor{MyGreen}{RGB}{10,120,90}

\usepackage{subcaption,setspace,enumerate}
\usepackage{graphicx,multicol}
\usepackage[normalem]{ulem}
\usepackage{soul}
\usepackage{adjustbox}

\usepackage{tabu}
\usepackage{makecell,multirow}

\usepackage{titlesec}
  \titleformat*{\section}{\sc \centering \Large}
  \titleformat*{\subsection}{\sc \centering \large}
  \titleformat*{\subsubsection}{\sc \large}
  \titlelabel{\thetitle. }
  \titlespacing*{\section}
  {0pt}{1.5ex plus 0.3ex minus .1ex}{1.5ex plus .1ex}
  \titlespacing*{\subsection}
  {0pt}{1.5ex plus 0.3ex minus .1ex}{1.5ex plus .1ex}
  \titlespacing*{\subsubsection}
  {0pt}{0.75ex plus 0.15ex minus .05ex}{0.75ex plus 0.5ex}

\definecolor{UBCblue}{rgb}{0.0, 0.25, 0.75}
\usepackage{float}
\usepackage[
            colorlinks = true,
            linkcolor  = black,
            citecolor  = black,
            urlcolor   = black,
            hyperfootnotes = false,
            hypertexnames = false,
            ]{hyperref}

\newcommand{\sep}[0]{ \; |\; }

\newcommand{\supp}[0]{ \text{{\normalfont supp}\;}}

\newcommand{\prob}[0]{ \text{{\normalfont Pr}}}

\usepackage[
            justification = justified,
            format        = plain,
            labelfont     = {bf,it},
            textfont      = it,
            belowskip     = 22.5pt,
            labelsep      = period,
            ]{caption}

\usepackage[
            authordate,
            backend      = biber,
            autocite     = inline, 
            noibid,
            maxcitenames = 3,
            doi          = false,
            isbn         = false,
            url          = true,
            eprint       = false,
            uniquename   = false,
            ]{biblatex-chicago}
\AtEveryBibitem{
  \ifentrytype{article}{\clearfield{url}}{}
  \ifentrytype{book}{\clearfield{url}}{}
  \clearlist{address}
  \clearfield{eprint}
  \clearfield{language}
  \clearfield{isbn}
  \clearfield{issn}
  \clearlist{location}
  \clearfield{series}
  \clearfield{urlyear}
  \clearfield{urlmonth}
  \clearfield{langid}
  \clearfield{urlday}
}

\renewbibmacro{in:}{}
\DeclareDelimFormat[cbx@textcite]{nameyeardelim}{\addspace}

\DefineBibliographyStrings{english}{
    backrefpage = {\lowercase{p}\adddot },
    backrefpages = {\lowercase{p}p\adddot }
}

\usepackage{tikz}
\usetikzlibrary{automata,patterns,intersections,arrows,decorations.pathreplacing,decorations.pathmorphing,positioning,arrows.meta,calc,decorations.markings,shapes.misc,matrix,shapes,fit,tikzmark,trees,calc,shapes,backgrounds, bending, babel}
\usepackage{pgfplots}
\pgfplotsset{compat = 1.16}

\usepackage{relsize}
\usepackage{scalefnt}

\usepackage{istgame}

\pgfdeclarepatternformonly{north east lines wide}%
   {\pgfqpoint{-1pt}{-1pt}}%
   {\pgfqpoint{10pt}{10pt}}%
   {\pgfqpoint{9pt}{9pt}}%
   {
        \pgfsetlinewidth{0.9pt}
        \pgfpathmoveto{\pgfqpoint{0pt}{0pt}}
        \pgfpathlineto{\pgfqpoint{9.1pt}{9.1pt}}
        \pgfusepath{stroke}
    }

\usepackage[
            margin = 1.10in, 
            top    = 1.10in,
            bottom = 1.10in,
            ]{geometry}
\usepackage[flushmargin]{footmisc}

\usepackage[strict]{changepage}

\newsavebox{\twosubbox}

\makeatletter
\def\th@plain{%
      \thm@notefont{}
      \itshape 
}
\def\th@definition{%
      \thm@notefont{}
      \normalfont 
}
\makeatother

\newtheorem{proposition}{\sc Proposition}

\newtheorem{remark}{\sc Remark}

\newtheorem{lemma}{\sc Lemma}
\newtheorem{lemma-app}{\sc Lemma}[section]
\newtheorem{assumption}{\sc Assumption}
\newtheorem{corollary}{\sc Corollary}

\newtheorem{definition-app}{\sc Definition}[section]

\theoremstyle{definition}

\usepackage{thmtools, thm-restate}

\usepackage{datetime}
\newdateformat{monthyeardate}{%
  \monthname[\THEMONTH] \THEYEAR}

\usepackage{afterpage}
\usepackage{xpatch}

\providecommand{\proofnamefont}{\itshape}
\xpatchcmd{\proof}{\itshape}{\normalfont\proofnamefont}{}{}
\renewcommand{\proofnamefont}{\bfseries}

\usepackage{enumitem}

\setlist[enumerate, 1]{
      itemsep   = 0.05em, 
      parsep    = 0.1em, 
      partopsep = 0.1em, 
      topsep    = 0.1em, 
      }

\setlist[itemize, 1]{
      itemsep   = 0.05em, 
      parsep    = 0.1em, 
      partopsep = 0.1em, 
      topsep    = 0.1em, 
      label     = $\bullet$,
      }

\setlist[itemize, 2]{
      itemsep   = 0.15em, 
      parsep    = 0.4em, 
      partopsep = 0.45em,
      label     = --,
      }

\setlist[itemize, 3]{
      itemsep   = 0.1em, 
      parsep    = 0.4em, 
      partopsep = 0.45em,
      label     = $\triangleright$
      }

\def\addlegendimage{\csname pgfplots@addlegendimage\endcsname}

\usepackage{nameref}
\usepackage[capitalise,noabbrev,nameinlink]{cleveref}
\crefname{table}{Table}{tables}
\crefname{equation}{Equation}{equations}
\crefname{lemma-app}{Lemma}{lemmas}
\crefname{claim}{Claim}{claims}
\crefname{problem}{Problem}{problems}
\crefname{example}{Example}{examples}
\crefname{assumption}{Assumption}{Assumptions}
\crefname{definition}{Definition}{Definition}
\crefname{definition-app}{Definition}{Definition}

\usepackage{comment}
\usepackage[textsize=scriptsize,textwidth=1in]{todonotes}

\title{Revealed and Concealed Repression:\\ Theory and Measurement}
\author{
\begin{tabular}{c @{\hspace{4em}} c}
Maria Titova\thanks{Department of Economics, Vanderbilt University.\\
E-mail: \texttt{motitova@gmail.com}} &
Nathan Canen\thanks{Department of Economics, University of Warwick and CEPR.\\
E-mail: \texttt{Nathan.Canen@warwick.ac.uk}} \\[1em]
Emily Hencken Ritter\thanks{Department of Political Science, Vanderbilt University.\\
E-mail: \texttt{emily.h.ritter@vanderbilt.edu}} &
Mehdi Shadmehr\thanks{Department of Public Policy and Department of Economics, UNC Chapel Hill.\\
E-mail: \texttt{mshadmeh@gmail.com}}
\end{tabular}
}
\date{}

\begin{document}

\maketitle

\thispagestyle{empty}
\begin{abstract}
Regimes routinely conceal acts of repression. We show that observed repression may be negatively correlated with total repression, consisting of both revealed and concealed acts. This distortion can generate perverse effects for policy interventions designed to reduce repression and complicates inference about the causes and consequences of repression. We develop a model in which regimes choose whether to conceal repression and activists decide whether to challenge the regime. We identify two measurement problems—one due to concealment and one to deterrence. We construct indices of repression that account for these problems and show how these indices can be expressed in terms of observable variables by leveraging equilibrium relationships. We then propose an empirical strategy to estimate these indices. As a proof of concept, we apply this approach to Russia, estimating repression indices at a monthly frequency for 2020–2025.

\medskip

\vspace{.3in}


\vspace{.1in}

\noindent \textbf{Word count:} 10,998 

\end{abstract}

\clearpage
\setcounter{page}{1}

\doublespacing

\section{Introduction}

Regimes routinely conceal acts of repression \autocite{guriev_spin_2022}. For example, the Iranian government has tried to hide the 1988 prison executions \autocite{Abrahamian-1999, Amnesty1988Massacre}, and the Chinese government has similarly tried to conceal the 1989 Tiananmen Square massacre \autocite{Reporters-2008}. Regimes remove social media posts about government arrests \autocite{king_how_2013}, punish dissidents for fabricated nonpolitical crimes when they could punish them for actual political offenses \autocite{pan}, and employ tactics such as stealth torture that are difficult to detect \autocite{rejali_torture_2007,conrad_torture_2018}. For example, during the Dirty War, the Brazilian military regime instructed its torturers to ``press hard without leaving marks'' \autocite[p. 180]{Dassin-1998}. Regimes also sometimes reveal acts of repression. Examples include the Saudi regime's mass execution of Shi`i dissidents \autocite{HRW-2022}, the Iranian regime's execution of protesters in the aftermath of a public challenge in 2022 \autocite{Amnesty-2023}, and the Chinese regime's execution of Uyghur activists \autocite{Guardian-2009}. Each case was reported, informing the public about the use of violence against political opponents. 

Because regimes sometimes conceal repression, observed repression is generally a distorted measure of total repression and its trend. Total repression, consisting of both concealed and revealed repression, can be significantly higher than observed repression alone. Importantly, the trends of total and observed repression can even move in \textit{opposite} directions. If a government shifts from revealed to concealed repression, observed repression may decline even as total repression increases. Consequently, policy interventions based on observed repression and its trend can be misguided and ineffective. The international community spends significant resources, in the form of sanctions, aid, or soft power, to mitigate repression.\footnote{Examples of U.S. legislation include the Global Magnitsky Human Rights Accountability Act of 2016, Foreign Assistance Act of 1961, section 502B, as amended, and the Leahy laws. Examples of international organizations resources include the World Bank's Human Rights trust fund (\url{https://www.worldbank.org/en/programs/humanrights}) and the European Union's Human Rights and Democracy Thematic Programme, and its predecessor, the European Instrument for Democracy and Human Rights.} However, governments and international organizations commonly use measures of observed repression (e.g., Human Rights Watch reports, Freedom House indices, and UN Universal Periodic Reports) to inform policies aimed at mitigating human rights abuses \autocite{lebovic_cost_2009, AndirinEtAl,poe1992human}.\footnote{\textcite{lebovic_cost_2009} find the World Bank reduces aid to those countries shamed by the UN Human Rights Commission/Council for rights violations. \textcite[p.~7]{AndirinEtAl} note that ``[T]he US Millennium Challenge Corporation incorporates Freedom House indices into its criteria to determine a country's eligibility for assistance \autocite{mcc_indicators_2020}. Canada’s Country Indicators for Foreign Policy project
integrates Freedom House indicators into data aimed at providing guidance to development-agency staff \autocite{carment-2010}.
The Open Government Partnership Global Report cites Freedom House data in the context of identifying potential areas
for future work and improvement \autocite[pp. 72, 78, 96]{ogp_globalreport_2019}.''}  
Regimes’ strategic concealment of repression also poses challenges for studies of the causes and consequences of repression \autocite{moore_repression_1998,moore_repression_2000,davenport2007state,earl_political_2011,guriev_spin_2022,HassanMattinglyNugent2022}, as researchers must identify causal effects in the presence of unobserved repression.

We study the measurement of repression in strategic environments in which regimes choose both whether to repress and whether to conceal repression, while dissidents choose whether to challenge the regime. We develop a model to identify the measurement problems arising from the concealment of repression and from the deterrence of dissidents, develop indices of repression that account for these problems, and show how these indices can be expressed in terms of observable variables by leveraging equilibrium relationships. We then propose an empirical approach to estimate these indices. As a proof of concept, we apply this approach to Russia, estimating the proposed repression indices for the period 2020--2025 and comparing the resulting estimates with existing measures.

In our model, dissidents decide whether to organize to demand changes to the status quo. These dissidents include activists, opposition members, and  ``vanguards'' in a movement \autocite{tarrow2011power,shadmehr2019vanguards}; we refer to them collectively as \textit{activists}. If activists organize, the regime decides whether to concede or repress, and whether to reveal or conceal repression at a cost. The public observes concessions, and it observes repression if it is revealed. The public then decides whether to protest. If the public protests when the activists have organized, the protest succeeds and the activists' demands are implemented. Otherwise, the protest fizzles and the status quo remains, reflecting the role of leadership and organizational resources in sustaining social movements \autocite{tilly1978mobilization, tarrow2011power,davenport_how_2015, sullivan_political_2016}.

The public does not take the desirability of change, the organizational capacity of the opposition, or the legitimacy of state coercion as given.\footnote{As in \textcite{shadmehr_international_2022}, the public is uncertain whether the activists' success will benefit or harm them, reflecting the uncertainties involved in major policy changes. 
The public expects the state to use coercion to protect it when necessary \autocite{almond1956comparative, mansbridge2012importance, mansbridge2014what, opp1990repression}, and therefore does not universally view coercion as illegitimate. Moreover, public opinion about activists, protesters, opposition groups, and the state itself  shifts in response to observed interactions between them \autocite{
chen2025scar,hager_does_2022, chenoweth2011why, chenoweth2023violence, wasow_agenda_2020, tertytchnaya_this_2023}. As captured in the phrase ``the revolution devours its children,'' regime changes can produce outcomes far worse than anticipated \autocite{shadmehr_collective_2011}.} 
Instead, the public makes inferences about these features. Revealed repression---or its absence---provides the public with information about two key aspects: (1) whether there are organized activists and a mobilization is underway, and (2) the beneficial or harmful nature of the activists’ demands and their success. In making its decision, the regime weighs the costs of concession, concealment, and the potential spread of mobilization from activists to the general public. The public, in turn, weighs the downside risk of protest if the activists’ success would worsen the status quo, the upside if their success would improve it, the likelihood of success, and the direct costs of protesting. Our model has two novel ingredients: the state's decision to conceal repression, and the public's inference about the presence of organized activists; because the government may conceal repression, the public remains uncertain about whether activists have organized unless it observes repression or concession.

We identify two problems in measuring repression: those arising from the concealment of repression and those arising from the deterrence of activists.
We illustrate the first by showing how lower concealment costs can increase total repression but reduce revealed repression, creating the appearance that the regime is becoming more tolerant. When concealment is easier, the regime substitutes from revealed to concealed repression and finds repression more attractive overall. Conversely, as concealment becomes more costly, total repression may decline even as revealed repression increases, making the regime appear less tolerant. Similar patterns arise with changes in protest costs. In sum, observed and total repression can move in opposite directions in response to environmental changes. Thus, measures that rely on observable repression not only overlook concealed acts but may also lead to incorrect inferences about trends in total repression. We provide empirical support for this logic by documenting a negative correlation between revealed and concealed repression under the Brazilian military regime, using data from the Brazilian National Truth Commission’s final report \autocite{memory}.

The measurement problem due to deterrence arises because repression can discourage activists from challenging the state in the first place. When the anticipation of repression deters activism, the state need not employ repression. Measures that rely on incidences of repression—even absent concealment—therefore capture only materialized repression and do not account for its deterrent effect \autocite{ritter-policy-2014,ritter_preventing_2016}.

We introduce two indices of repression that account for both concealed repression and the deterrent effect of repression. One captures repression against all activists, while the other captures repression against activists whose success the public views as beneficial.\footnote{These indices correspond to two normative perspectives: one views the use of coercive force against all activists as illegitimate, while the other considers it the duty of the state to protect the public against organized activists whose success harms the public.} Each index measures total repression, encompassing both revealed and concealed repression. We show how total repression can be recovered from observed repression by leveraging equilibrium relationships between observable and unobservable variables. In particular, we express total repression as a function of observed repression and the public’s beliefs about the opposition, which can be measured through surveys.

We propose an empirical approach to estimate these indices using data on observed repression and public opinion surveys about activists and establish the consistency of the estimators. As a proof of concept, we estimate the indices of repression for Russia from 2020 to 2025 at a monthly frequency. We compare our measures with those from Freedom House, Varieties of Democracy, and the Political Terror Scale. Our measure of revealed repression follows a similar upward trend to the Freedom House and Varieties of Democracy measures, suggesting that the regime has become more repressive. By contrast, our measure of total repression remains close to its maximum throughout the period, indicating that lower levels of revealed repression in earlier years may reflect a greater prevalence of concealed repression, particularly before the invasion of Ukraine. Our measure of total repression against activists whose success the public views as beneficial declines over this period, suggesting that the public increasingly viewed state actions against activists as legitimate uses of coercion. 
We conclude by discussing the feasibility of obtaining reliable public opinion data in authoritarian regimes beyond Russia.

Our paper contributes to the study of measurement problems in political economy when the variable of interest is partially unobserved under conditions that are \textit{a priori} unknown. One example arises when actors strategically opt out of interaction. Heckman-type selection models offer procedures to reduce bias by statistically modeling the selection process \autocite{wooldridge2010econometric}. \textcite{ritter-policy-2014} and \textcite{conrad_contentious_2019} adopt this approach to study the determinants of repression. In contrast, \textcite{AndirinEtAl} take a structural approach to measuring unobserved repression,\footnote{This approach has been used in ideal point estimation \autocite{mccarty2006polarized}, bargaining models \autocite{signorino1999,DiermeierEtAl-2003}, and third-party intervention in conflicts \autocite{GibiliscoMontero2022}.} mapping unobserved preventive repression to a model parameter that is then estimated. In the same spirit, we use equilibrium relationships to recover unobserved repression from observable variables.

Recovering unobserved total repression from observable parameters is especially important when observed and unobserved repression do not have a strong positive correlation across space and time. In such cases, relying solely on observed repression can lead to perverse policy interventions that produce effects opposite to those intended. Specifically, interventions may mistakenly be deemed successful when they are not, and vice versa, and inadvertently reward regimes that have increased total repression and penalize those that have reduced it. The problem of identifying the true trend of repression over time has been the subject of recent academic debates \autocite{Cingranelli-Filippov-2018, Fariss-2019, Cope-Crabtree-Fariss-2020}. \textcite{fariss_respect_2014} points out that time trends in repression measures may be misleading because the standards used by experts and reporters to classify actions as repression can change over time. In contrast, the problem we focus on concerns the regime’s strategic decision to conceal repression.
However, there is an overlap: If the ability of observers to detect repression improves more than the ability of the regime to conceal it, this is equivalent to an increase in concealment costs in our model. In that case, observed (revealed) repression may rise even as total repression falls, potentially leading to misguided and damaging policy interventions.

Methodologically, our theoretical model falls into an unexplored class of disclosure games with a type-dependent outside option for the sender. Our model builds on the signaling framework of \cite{shadmehr_international_2022} by enabling the regime to conceal repression. The option to conceal transforms the model into a disclosure game. Absent concession and concealment costs, our model is one of disclosure of verifiable information \autocite{Hart-et-al-2017,GieczewskiTitova}, and we use the well-known insight that no news may reflect that the sender is potentially uninformed \autocite{Dye-1985,shadmehr-bernhardt-2015, Besley-Prat-2006,Verrecchia-2001}. However, existing models do not feature the type-dependent outside option (concession), which is important for our substantive application.

We next review common approaches to measuring repression and compare them with our proposed approach. We then present a model of revealed and concealed repression and derive its implications for measurement, identifying problems arising from concealment and deterrence. In the following section, we introduce our indices of repression, show how total repression can be expressed in terms of revealed repression and public opinion about the opposition, propose an empirical strategy to estimate the indices, and apply it to Russia from 2020 to 2025.

\section{Common Approaches to Measuring Repression}\label{sec:existing-measures}

Existing approaches to measuring  repression typically fall into two categories: event counts and indices based on expert opinion.\footnote{See, however, \textcite{AndirinEtAl}, who use a structural estimation approach to develop a measure of preventive repression as the maximum of the support of the distribution of equilibrium protest probabilities in a given time and location.}
In the first approach, incidents of observed repression (e.g., beating protesters) are recorded based on sources such as news wires or social media in datasets such as the Social Conflict Analysis Database (SCAD) \autocite{scad}, the Armed Conflict Location and Event Data (ACLED) \autocite{acled}, and the Global Data on Events, Language, and Tone (GDELT) \autocite{gdelt}, and are then aggregated into a measure of repression.
In the second approach, expert opinions are collected in the form of human rights reports or surveys and are then mapped into a measure of repression. For example, Amnesty International, Human Rights Watch, and the US State Department publish annual reports that scholars use to construct measures of repression, e.g., as in the Political Terror Scale \autocite{wood_political_2010,pts} and the CIRIGHTS Project \autocite{mark_cirights_2023}. \cite{freedom_house_freedom_2025}, Varieties of Democracy \autocite{coppedge_methodology_2019}, and the Human Rights Measurement Initiative \autocite{brook_human_2020} each surveys experts and then maps the results into indices.

When  states devote substantial resources to concealing repression, some forms of repression remain hidden despite all efforts to uncover them. Even when the presence of a form of repression—such as killing or disappearance—is known, its extent and details may remain uncertain. For example, experts knew that the Guatemalan National Police engaged in forced disappearances of dissidents during the civil war of the 1980s, and there were allegations of other repressive acts, including massacres, which the government routinely concealed. A fact-finding report issued after the end of the war writes, ``For many years they [the people of Guatemala] were unable to share their experience, to reveal what had happened, and to accuse those responsible'' \autocite[xxxi]{archdiocese_of_guatemala_guatemala_1999}.
The details and extent of the repression were revealed only years later by the UN-mandated Historical Clarification Commission and through classified national police files discovered and revealed by a human rights organization \autocite[15--16]{sullivan_political_2016}. Truth commissions after other instances of regime change—for example, in Argentina \autocite{osorio_argentina_2022}, South Africa \autocite{south_african_truth_and_reconciliation_commission_truth_1999}, and Brazil \autocite{memory,memory2,memory3}—similarly revealed numerous previously concealed acts of repression. Even in democratic countries such as the U.S. and Japan, discriminatory police surveillance of potential dissidents has been concealed from experts and the public at the time it occurred and revealed only after the fact when classified documents were leaked \autocite{police_2014,repeta_spying_2016}.

The extent to which human rights organizations can uncover concealed repression depends on their resources and priorities. The number and background of experts vary considerably across institutions that produce repression indices. For example, Freedom House ``involved 136 analysts and around 45 advisers'' across all countries in 2025 \autocite[p.~2]{freedom_house_freedom_2025}. According to V-Dem, it has ``more than 4,000 Country Experts\dots providing expert information via our online surveys\dots Experts are usually academics or professionals with specialist and evidenced knowledge\dots 
The quality and impartiality of the data is highly dependent on the Country Experts'' \autocite{varieties_of_democracy_v-dem_2026}. Human Rights Measurement Initiative relies on ``human rights practitioners who are actively monitoring the civil and political rights situation in each country\ldots In most cases survey respondents are located within the country'', but ``for some more closed countries,'' it uses ``a higher proportion of respondents who are based outside of the country in question'' \autocite[719--720]{clay_using_2020}. This difficulty of identifying experts who will report from closed, autocratic countries can prevent data production; HRMI, for instance, currently does not have ratings for Russia or Iran.\footnote{For an updated list of countries covered by the Human Rights Measurement Initiative, see \url{https://
humanrightsmeasurement.org/country-coverage}.} Since it does ``not have the capacity to vet all potential survey respondents,'' HRMI ``work[s] through trusted partners, and a network of HRMI Ambassadors, who help to connect\dots to potential survey respondents''  \autocite[719--720]{clay_using_2020}.

The literature highlights potential biases in indices that rely on expert ratings and reports. For example, ratings based on reported allegations do not capture unreported incidents \autocite[p.~430]{conrad_disaggregating_2013}. For more subjective ratings, the availability of information and its interpretation may vary substantially across experts and over time \autocite{clark_information_2013,fariss_respect_2014,little_measuring_2024}. Rating institutions may exhibit political bias  \autocite{steiner_comparing_2016,bush_politics_2017} or focus unevenly on particular forms of repression, such as imprisonment and torture \autocite{kingsbury_international_2015,clay_using_2020}. Although human rights organizations make significant efforts to address concerns, it remains difficult to assess the knowledge, bias, or selection process of experts whose identities are routinely kept confidential for their protection. For example, even a highly ethical expert may decide to exaggerate the extent or credibility of available information, or to downplay marginal human rights progress, in order to induce intervention by decision-makers who might otherwise be reluctant to spend limited resources on a particular country.\footnote{These problems are not unique to human rights experts. In the context of election fraud, the literature identifies the biases of monitors, who may adjust their reports to avoid electoral violence \autocite{Carothers1997,Kelley2012,LongEtAl2013,luo2018election}. It also highlights governments’ strategic responses, including substitution across different tactics and locations \autocite{SimpserDonno2012,Ichino2012,hyde2011pseudo}, as well as strategies of creating or inviting additional observers to dilute the impact of negative reports \autocite{Carothers1997,hyde2011pseudo,Kelley2012,DaxeckerSchneide2014,Norris2017,MorrisonEtAl2024}.}

These critiques underscore the value of alternative approaches to measuring repression. Scholars and institutions continue to develop methods that reduce bias and improve the detection and credible reporting of repression. Our approach, however, does not attempt to directly detect or measure concealed repression. Instead, we leverage equilibrium relationships to recover concealed repression from observed revealed repression and other observable variables. The regime’s decisions to conceal or reveal repression are interdependent, as it may substitute between these strategies. If, conditional on observable variables, there is a one-to-one mapping between the likelihoods of concealed and revealed repression, we can recover one from the other. We identify conditions under which this mapping holds conditional on the public’s beliefs about activists and opposition groups. Factors that differentially affect the regime’s incentives to reveal or conceal repression are then captured through their effects on public beliefs.

The approach proposed here is less susceptible to regimes’ strategic adaptations aimed at avoiding detection or substituting toward less observable forms of repression, because the researcher does not attempt to measure unobserved repression directly. However, this approach also has limitations. It relies on the consistency of beliefs and strategies that underlies much of modern game theory; if beliefs become sufficiently decoupled from strategies, the approach breaks down. We also assume standard Bayesian updating. While alternative updating rules would change the exact formulas, the approach can be adapted accordingly. Finally, we implement the approach within a particular model and demonstrate its robustness to some classes of modifications in \cref{sec:robustness}, leaving its broader robustness to alternative modeling choices for future research.

\section{Model of Revealed and Concealed Repression}\label{section:model}

We begin by outlining the model informally before turning to its formal exposition. Consider a society consisting of a regime, the general public, and activists. 
To maintain the status quo, the regime requires some degree of support from the general public. 
If activists challenge the regime and the public supports them, the status quo is overturned.

Activists differ in the tactics and policies they pursue. Some pursue actions and policies that are beneficial for the public (e.g., Solidarity in Poland); we refer to them as ``good'' activists. Others pursue actions and policies that are harmful to the public (e.g., the Mojahedin in Iran); we refer to them as ``bad'' activists. This distinction encompasses both the goals activists pursue and the tactics they employ, including peaceful mobilization, targeted violence against security forces, or indiscriminate violence. The regime and the public have a conflict of interest: The regime prefers to avoid challengers and maintain the status quo over any change.

First, activists decide whether to organize and challenge the regime. At this stage, activists face uncertainty about how they will ultimately be perceived by the public, reflecting, for example, imperfect knowledge of public preferences and unpredictable incidents and difficult decisions that arise during mobilization, such as fundraising through illegal means, justified or unjustified violence, or damage to public or private property. 
With probability $q$, the activists’ success benefits the public; with probability $1-q$, it harms the public. If activists organize, the regime can either concede to their demands or repress them, and either conceal or reveal that repression.

The public then decides whether to protest against the regime. A protest succeeds if and only if activists have organized and the regime has not conceded to their demands. If protest succeeds, the activists’ demands are implemented and the regime incurs a cost. Importantly, the public is uncertain both about whether activists have organized and about whether their success would be beneficial or harmful \autocite{shadmehr_international_2022, shadmehr_collective_2011}. When the public observes concessions, it infers that protest would be futile, either because activists are satisfied or because they have been co-opted, and therefore does not protest. When the public observes repression, it knows that protest would succeed if undertaken, but must decide whether to side with potentially harmful activists against the regime. By contrast, when the public observes neither concession nor repression, it remains uncertain whether activists organized, or whether the regime repressed organized activists and concealed that repression. These uncertainties and the resulting trade-offs are central to our analysis.

Formally, there is a regime, a public, and an activist. First, the activist decides whether to organize, making demands on the regime.  If the activist organizes, nature draws the activist’s alignment with the public. With probability $q \in (0,1)$, the activist’s actions and demands align with the public’s preferences (i.e., the activist is ``good'', denoted $G$); with probability $1-q$, they do not (i.e., the activist is ``bad'', denoted $B$). The regime observes whether the activist has organized and is good or bad, and then decides whether to concede to the activist, to repress-and-reveal (revealed repression), or to repress-and-conceal (concealed repression). Repress-and-conceal costs the regime $c$, drawn from a distribution $H$ with full support on $[0,\overline{c}]$. If the activist does not organize (denoted by $N$), the regime does not have a decision. Let $\theta \in \{G,B,N\}$ denote the state of the world regarding the activist: whether the activist has organized, and if so, whether the activist is good or bad. We can think of $(\theta,c)\in \{G,B,N\}\times [0,\overline{c}]$ as the regime's two-dimensional type.

The term repress-and-conceal refers to the combination of carrying out repression and concealing it from the public. It takes two broad forms. First, some repression tactics are inherently less observable to the public, and regimes may select those tactics when they want to conceal repression \autocite{rejali_torture_2007,conrad_torture_2018}. These include stealth torture \autocite{rejali_torture_2007}, covert surveillance \autocite{nalepa_authoritarian_2022}, the weakening of activist organizations \autocite{davenport_how_2015,nalepa_authoritarian_2022,sullivan_political_2016}, subtle intimidation, and the use of fabricated non-political charges \autocite{pan}. Second, regimes actively suppress information about repression through media control. This includes co-opting or threatening journalists \autocite{mazzaro_anti_2023, rwb_in_2023}, flooding social media with misinformation or irrelevant content to drown out news of repression, establishing state-controlled media, and controlling access to the internet and digital platforms \autocite{roberts_censored_2018}. 
The costs of concealment reflect the opportunity costs of the expertise, coordination, resources, and infrastructure required to hide repression. These costs underscore the high levels of sophistication, organization, and technology needed to carry out concealed repression \autocite{guriev_spin_2022, greitens2016dictators, Berajaetal2023}. In fact, economic windfalls to regimes have been shown to increase the prevalence of less observable forms of repression \autocite{AndirinEtAl}.

Next, the public observes concession if it occurred, and it observes repression if and only if the regime chose to repress-and-reveal. Otherwise, the public observes no news, so it cannot tell whether no activist has challenged the regime or the regime has chosen to repress-and-conceal. The public then decides whether to protest at a cost $\rho$ drawn from a distribution $F$ with full support on $[0,\overline \rho]$. The protest cost $\rho$ is the public's private information.\footnote{Results are identical if $\rho$ is drawn and the public is informed after the regime's decision.} The protest succeeds if and only if there is an organized activist (that is, $\theta\in\{G,B\}$) and the regime has not conceded. If the protest succeeds, the activist's demands are implemented, and the regime pays a cost.
The payoffs are realized, and the game ends.

The activist's payoff, if it does not organize, is normalized to $0$. The activist receives a benefit $b>0$ if the activist's demands are implemented (through concession or successful protest) and pays a cost $r\geq 0$ if the regime represses the activist. This cost-to-benefit ratio $r/b$ is drawn from a distribution $\Gamma$ with full support on $[0,\infty)$. This cost-to-benefit ratio $r/b$ is the activist's private information.\footnote{Our modeling choices reflect that the activist decision to organize is not informative about whether their tactics and demands will ultimately be beneficial or harmful for the public.} The regime's payoff following a successful protest by the public is normalized to $0$, and its payoff from the status quo is normalized to $1$. Its payoff from conceding to an activist (and avoiding a successful protest) is $1-\alpha_G$ when the activist is good, and $1-\alpha_B$ when the activist is bad, with $\alpha_G, \alpha_B \in(0,1)$. The regime pays the concealment cost $c$ if it decides to repress-and-conceal.  The public's payoff from the status quo is normalized to $0$. Its payoffs from the success of the good and the bad activists are $\beta_G$ and $\beta_B$, respectively, with $\beta_G>0>\beta_B$. The public pays the cost $\rho$ if it protests. The random variables $r/b,\ c$, and $\rho$ are independent from each other, and their corresponding CDFs $\Gamma,\ H$, and $F$ are continuously differentiable. 

\cref{sec:robustness} generalizes the model by introducing direct costs of revealed repression for the regime and allowing for imperfect concealment, such that the regime’s attempt to conceal repression may fail.

We focus on perfect Bayesian equilibria in which repress-and-reveal occurs with strictly positive probability, so that the model’s implications remain consistent with empirical patterns.\footnote{\cref{lemma:regime-reveals-under-D1} in \cref{sec:appendix-proofs} shows that these are precisely the perfect Bayesian equilibria that satisfy the D1 refinement---see \textcite{crawford2011theory,BANKS19941} for experimental analyses of equilibrium refinements.} The activist's strategy is a mapping $\psi: [0,\infty) \to [0,1]$ from its cost-to-benefit ratio to a probability of organizing. The regime's strategy is a mapping $$\sigma: \{ G,B \} \times [0,\overline{c}] \to \Delta \{ \text{concede}, \text{repress-and-reveal}, \text{repress-and-conceal} \}$$ from the regime's type $(\theta,c)$ to a probability distribution over the regime's actions. The public's strategy is a mapping $\phi_\omega: [0,\overline \rho]\rightarrow [0,1]$ from its protest costs $\rho$ to a probability of protesting upon observing $\omega \in \{ RR,NN \}$, where $RR$ indicates revealed repression and $NN$ indicates observing neither repression nor concession, i.e., no news. 

We maintain the following assumption throughout the paper. For clarity, we restate the support properties of the distributions. 

\begin{assumption}\label{A1} Let $\beta^e:= q\beta_G +(1-q)\beta_B$ 
  be the public's expected policy payoff from successful protest under the prior. Suppose that
    \begin{enumerate}
        \item $\alpha_G<\alpha_B$; \label{A1-alphaG<alphaB}
        \item $\alpha_G<F(\beta^e)$; \label{A1-betaE-GalphaG}
        \item $\alpha_G>F(\beta_B)$; \label{A1-betaB-GalphaG}
        \item $\supp H = [0,\overline{c}]$, $H(0) = 0$, $\overline{c} > \alpha_B$; \label{A1-suppH}
        \item $\supp F = [0,\overline{\rho}]$ and $F(0) = 0$; \label{A1-suppF} 
        \item $\supp \Gamma = [0,+\infty)$ and $\Gamma(0) = 0$. \label{A1-suppGamma} 
    \end{enumerate}    
\end{assumption}
The first part, $\alpha_G<\alpha_B$, is substantive, reflecting that the regime dislikes conceding to organized bad activists more than conceding to organized good activists; the conflict of interest between the regime and the public is mild.  Consequently, after observing repression, the public remains uncertain about whether $\theta = G$ or $B$ in equilibrium---see \cref{prop:equilibrium}, Part 3. 
The remaining parts ensure that available actions occur with  positive probability in equilibrium. In particular, the equilibrium probabilities that the activist organizes, the regime engages in repress-and-conceal, repress-and-reveal, and concede, and the public protests absent news or after observing repression are all strictly positive and less than 1, so that the model’s implications remain consistent with empirical patterns.

\subsection{Analysis}\label{sec:analysis}

We begin analysis with the public's decision. Let $\mu = ( \mu(G), \mu(B), \mu(N) ) \in \Delta(\Theta)$ be the public's belief. We denote by $\mu_{RR}$ the public's posterior belief upon observing repression and by $\mu_{NN}$ its posterior belief when the public observes no news.
Given a belief $\mu$, the public protests if and only if its (posterior) expected payoff from protest exceeds its costs: $\mu(G) \beta_G + \mu(B) \beta_B -\rho\geq 0$. It follows that the public protests whenever direct protest costs are below a threshold that depends on the public's belief about the activist.

\begin{lemma}\label{lemma1:eqm-revolutions}
    In any equilibrium, the public protests if and only if $\rho\leq \widetilde\rho(\mu)$, where $\widetilde{\rho}(\mu):=\mu(G)\beta_G+\mu(B)\beta_B$ and $\mu \in \Delta(\Theta)$ is the public's belief.
\end{lemma}

All proofs are in \cref{sec:appendix-proofs}.  Similarly, the regime chooses to repress-and-conceal an activist whenever the concealment cost is below some threshold $\widetilde{c}_\theta$, for $\theta \in \{ G,B \}$. \cref{prop:equilibrium} shows that these thresholds are the same for good and bad activists in any equilibrium: $\widetilde{c}_G=\widetilde{c}_B=\widetilde{c}$. 

\newpage
\begin{proposition}\label{prop:equilibrium}
There exists an equilibrium.   In any equilibrium: 
    \begin{enumerate}
      \item The activist organizes whenever $r/b$ is below a threshold. This happens with a probability $\gamma \in (0,1)$, where $\gamma=\Gamma(\Phi(\gamma))$ and $\Phi(\gamma)$ is given by \eqref{eqn:Phi-of-gamma}.
        \item If $c < \widetilde{c}$,  the regime represses all organized activists and conceals that repression, where $\widetilde{c} \in (0,\alpha_G)$ is the unique solution to
    \begin{equation}\label{eqn:tilde-c}
            \frac{\gamma H(\widetilde{c})}{\gamma H(\widetilde{c}) + 1-\gamma}\ \beta^e  = F^{-1}(\alpha_G-\widetilde{c}).
    \end{equation}
    \item If $c > \widetilde{c}$, the regime publicly represses organized bad activists and represses organized good activists with probability less than 1. In particular, the  likelihood ratio of publicly repressing good versus bad organized activists is 
    $$\kappa := \frac{\int \sigma(\text{repress-and-reveal}\sep G,c) dH(c)}{\int \sigma(\text{repress-and-reveal}\sep B,c) dH(c)} = \frac{1-q}{q} \frac{F^{-1}(\alpha_G) - \beta_B}{\beta_G - F^{-1}(\alpha_G)} \in (0,1).$$
    \item When the public does not observe concession or repression, it believes that the presence of an organized opposition is less likely, but it does not update its beliefs about whether the activist is good or bad: $\mu_{NN}=(\gamma' q,\gamma' (1-q),1-\gamma')$, where $\gamma' := \frac{H(\widetilde{c})\gamma}{H(\widetilde{c})\gamma + 1-\gamma} < \gamma$.
\end{enumerate}    
\end{proposition}

It is intuitive that the activist organizes and makes demands on the regime with an endogenous, strictly positive probability less than $1$. \cref{prop:equilibrium} also shows that if the regime plans to conceal repression, it represses and conceals all activists, good and bad. The intuition hinges on the assumption that it is less costly for the regime to concede to good activists than to bad ones (\cref{A1}.\ref{A1-alphaG<alphaB}). Upon observing a regime action, the public responds in the same way regardless of whether the activist is good or bad ($\theta = G$ or $\theta = B$), because it does not observe that information. Together, these imply that the regime engages in repress-and-conceal of both good and bad activists equally. Suppose it does not. Because the regime's alternatives to repress-and-conceal are more attractive when activists are good, if the regime engages in repress-and-conceal less frequently for one type of activist, it is the good activists. In that case, whenever the regime does not repress-and-conceal good activists, it concedes to them. It follows that the regime never represses and reveals good activists. But then, following observed repression, the public infers that the activists are bad and therefore does not protest. Anticipating this, the regime prefers to repress-and-reveal good activists rather than concede to them. Therefore, the regime engages in repress-and-conceal of both good and bad activists equally.

An implication of \cref{prop:equilibrium} is that the public does not infer anything about the nature of the activists' demands following no news: The posterior likelihood ratio that the activists are good versus bad remains the same as the prior.
However, the absence of news of concession or repression can be informative about the presence of organized activists. Because activists only sometimes organize to challenge the regime ($\gamma<1$) and because the regime sometimes conceals repression $\widetilde{c}<\overline{c}$, following no news, the public updates its belief that it is less likely that activists have organized:
\begin{eqnarray*}\label{eqn:gamma'-gamma}
\Pr(\theta\in\{G,B\} \sep \text{no news}) = \frac{\gamma\ H(\widetilde{c})}{\gamma\ H(\widetilde{c})+1-\gamma}<\gamma.
\end{eqnarray*}

Upon observing repression, by contrast, the public learns that activists have organized and updates its belief about whether the organized activists are good or bad. In particular, 
let $q'$ denote the public's belief, upon observing repression, that the activists are good:
\begin{equation}
    q'=\prob(G \sep \text{revealed repression} ).\notag
\end{equation}
Since the regime observably represses bad organized activists more often (\cref{prop:equilibrium}, Part 3), the public updates negatively about the organized activists that are observably repressed: 
\begin{equation}\label{eqn:q'<q}
q'<q.    
\end{equation}
This implication follows from \cref{A1}.\ref{A1-alphaG<alphaB} and is consistent with the empirical results in \cref{sec:Russia} (see footnote~\ref{fn:qqprime}) and with the literature showing that survey respondents are less likely to approve of or express support for protesters who have experienced repression than for those who have not \autocite{tertytchnaya_this_2023,pechenkina_how_2019}.

\cref{prop:equilibrium} allows us to characterize the ex ante probabilities of revealed, concealed, and total repression, which we will use in subsequent sections to analyze the problem of measuring repression and to suggest indices of repression.
Upon observing a bad activist, the regime will repress-and-reveal with probability $1-H(\widetilde{c})$. 
Upon observing a good activist, the regime will repress-and-reveal with probability $\kappa( 1-H(\widetilde{c}) )$.
Therefore, the ex-ante probabilities are: 
\begin{align}
    \prob(\text{concealed repression}) &= \gamma H(\widetilde{c}) \label{eqn:concealed-repression} \\
    \prob(\text{revealed repression}) 
    &= \gamma (\kappa q + 1-q)(1-H(\widetilde{c})) \label{eqn:revealed-repression} \\
    \prob(\text{total repression}) &= \prob(\text{revealed repression}) + \prob(\text{concealed repression}), \label{eqn:total repression basic formula}
\end{align}
where $\gamma$, $\kappa$, $\widetilde{c}$ are equilibrium objects described in \cref{prop:equilibrium}.\footnote{A formal literature studies how regimes combine repression with information manipulation when the information is unrelated to repression \autocite{guriev_theory_2020,EgorovSonin2021,GitmezMolaviSonin2025,GehlbachLouEtAl2025,EgorovSonin2024}. Our focus is instead on the regime's concealment of information about repression itself. Interpreting $H(\tilde c)$ as the probability of censorship (of repression), equations \eqref{eqn:concealed-repression}-\eqref{eqn:total repression basic formula} imply $\prob(\text{total repression})=\gamma[(\kappa q + 1 - q) + q(1 - \kappa)\, H(\widetilde{c})]$, suggesting that repression and the censorship of repression may be complements in this setting.}

\begin{remark}[aligned regimes]\label{remark:aligned regime}
\setlength{\belowdisplayskip}{-10pt}
In our environment, the regime pays a cost  $\alpha_G>0$ by conceding to good activists, creating a conflict of interest with the public who benefits from such concessions. It is straightforward to see that, if $\alpha_G=0$, there is an equilibrium in which the regime concedes to organized good activists and publicly represses organized bad activists. Let $\prob_{\text{aligned regime}}(\text{total repression})$ be the probability of total repression in this equilibrium with a regime whose interests are aligned with the public. Then, 
\begin{equation}\label{eqn:aligned regime}
   \prob_{\text{aligned regime}}(\text{total repression})=\gamma(1-q)>0.
\end{equation}
\end{remark}
\noindent This probability is strictly positive, highlighting that even a regime whose interests are aligned with the public will engage in some level of repression of activists whom the public views as harmful. \cref{remark:aligned regime} highlights the connection between formalizations of repression and normative notions of the legitimate and illegitimate use of coercion by the state \autocite{mansbridge2012importance,mansbridge2014what,almond1956comparative,opp1990repression}. As part of what we term “total repression,” equation \eqref{eqn:aligned regime} counts the regime’s repression of (i.e., use of coercion against) activists whose demands are harmful to the public and those whose demands are beneficial. This is consistent with the idea that the use of coercive force to control opposition constitutes repression regardless of the opposition's identity or claim \autocite{earl_political_2011}. For example, in their review of the literature on repression, \textcite{earl_digital_2022} note that scholars ``may struggle with the idea that both (ultra)liberal and (ultra)conservative activists and movements can be repressed\dots The repression literature shows, however, that repression happens to both, although not necessarily in equal measure, using the same forms, or to the same ends'' (p.~12). 

In contrast, one might argue that some instances of coercion may be legitimate, in which the regime acts appropriately to protect the public from harmful activists and therefore should not contribute to a measure of repression. As \textcite{earl_digital_2022} note, ``for many [scholars], repression is what happens to the groups you favor, whereas lawful governance is what affects groups you dislike'' (p.~12). For example, if activists threaten innocent bystanders, one might argue that the state has a duty to use coercive force to stop them \autocite{edwards_violence_2021}. However, it can be difficult to obtain accurate information about specific incidents, as both the state and activists routinely deny wrongdoing and attribute blame to the other side. The public may also support the use of state coercion against actors perceived as security threats \autocite{conrad_threat_2018,aksoy_effect_2022}, as violent \autocite{lupu_violence_2019,edwards_violence_2021,naunov_effect_2025}, or as lawbreakers \autocite{tertytchnaya_this_2023}. Of course, this normative perspective does not account for divergences between conceptions of human rights and what the public—however defined—perceives as beneficial, nor for variation across time and place in public judgments about what constitutes beneficial or harmful demands (including their moral dimensions).

Recognizing the value tradeoffs involved across these perspectives, we take a middle ground by proposing two indices in \cref{sec:measuring repression}: one that captures repression of all activists, and another that focuses on repression of activists whom the public views as “legitimate.” We acknowledge that researchers or human rights organizations may wish to adopt alternative normative criteria, and we hope that the methodological approach developed here will prove useful for constructing indices aligned with alternative notions.

\subsection{Implications for Using Measures of Revealed Repression} \label{subsec:measurement problem}

\cref{prop:equilibrium} highlights two types of strategic decisions that can distort the measurement of repression. 
The first has to do with the regime's strategic decision to conceal ($\widetilde c>0$). Revealed acts of repression are observable, so their  probability can be empirically estimated. Concealed acts are not observable---at least until sufficient time has passed and significant investigations have been conducted. This creates a {\it measurement problem due to concealment}: if researchers or policymakers focus only on observable repression, they risk drawing misleading inferences about the overall prevalence of repression. 
The second has to do with the activists' strategic decision not to challenge the state in order to avoid repression. This creates a \textit{measurement problem due to deterrence}: the threat of repression deters the activist from organizing in the first place, reducing the need to exercise repression \autocite{ritter-policy-2014, ritter_preventing_2016}. 

\subsubsection{Measurement Problem Due to Concealment}

We begin with the first problem. Despite concealed repression, one may conjecture that revealed and total repression are strongly positively correlated. Then, while revealed repression is a lower bound on the overall repression, at least it captures the trends of total repression. However, this is not generally true. 

To illustrate the extent of the problem, we simulate the equilibrium probabilities of observable (revealed) repression and unobservable (total) repression as functions of concealment and protest costs. We assume that the distribution of concealment costs is $H = \text{Beta}(\lambda_H,1)$ and the distribution of protest costs is $G = \text{Beta}(\lambda_G,1)$, with $\lambda_H, \lambda_G > 0$. An increase in $\lambda_H$ or $\lambda_G$ raises the corresponding costs in the sense of first-order stochastic dominance.

\begin{figure}[t]
    \centering
    \begin{subfigure}[b]{0.49\textwidth}
        \centering
        \includegraphics[width=0.98\linewidth]{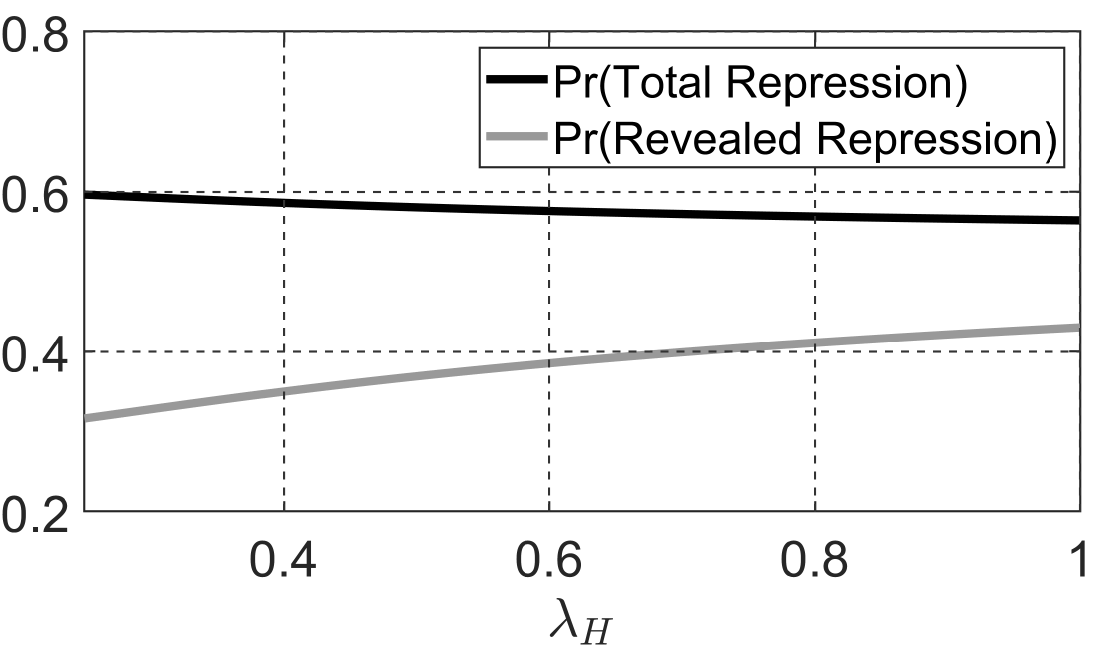}
        \label{fig:H-comp-stat}
    \end{subfigure}%
    ~ 
    \begin{subfigure}[b]{0.49\textwidth}
        \centering
        \includegraphics[width=0.98\linewidth]{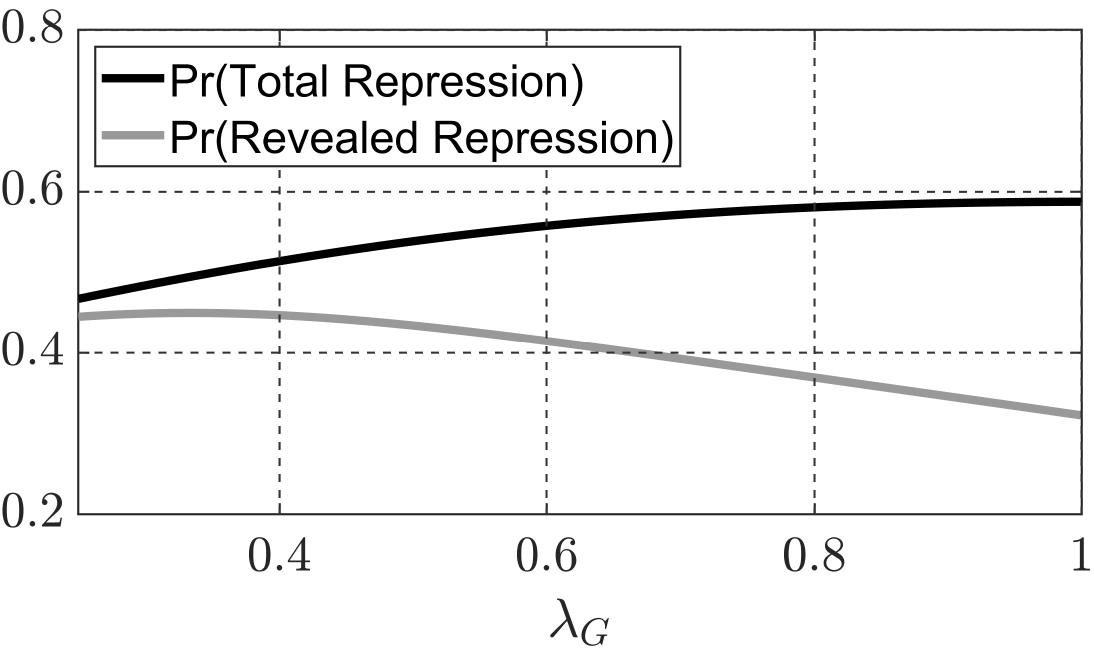}
        \label{fig:G-comp-stat}
    \end{subfigure}
    \caption{Negative correlation between total and revealed repression. Basing analysis or policy on the observable revealed repression can lead to perverse outcomes.  Left Panel: $F = Beta(0.8,1)$ and $H = Beta(\lambda_H,1)$; Right Panel: $G=Beta(\lambda_G,1)$ and $H = Beta(0.5,1)$. Parameters: $q=0.65$, $\Gamma = Exp(1.5)$ $\beta_G=2$, $\beta_B=-0.5$, $\alpha_G=0.7$, and $\alpha_B = 0.95$. 
    }
    \label{fig:HandG-comp-stats}
\end{figure}

 \cref{fig:HandG-comp-stats} depicts the simulations. Reductions in concealment costs increase unobservable total repression while reducing observable revealed repression. As a result, observers may mistakenly infer that the regime is becoming more tolerant. Conversely, when concealment costs increase, total repression can fall while revealed repression simultaneously increases, making it appear to observers that the regime is becoming less tolerant. These patterns are not special to concealment costs. As \cref{fig:HandG-comp-stats} illustrates, similar negative correlations and opposite trends arise in response to variations in protest costs. 

The underlying logic is that total repression consists of both revealed and concealed repression, and these two components may be negatively correlated, as regimes substitute between them when their relative costs change. We now provide empirical evidence for this negative correlation in the context of the Brazilian military regime from 1964 to 1979, prior to the regime’s gradual opening known as ``abertura'' \autocite{skidmore}.

The Brazilian military regime kidnapped, tortured, and killed political activists and routinely concealed these repressive acts. Victims’ bodies were secretly removed from military facilities and buried in clandestine graves or thrown into the ocean \autocite{memory,memory2,memory3}. The full extent of these killings became publicly known only decades later through extensive investigations conducted by Brazil’s National Truth Commission, which was established in 2011 and released its final report in 2014. As a measure of \emph{concealed repression}, we use annual counts of political killings reported in Volume~1 of the Commission’s final report \autocite[pp.~487--492]{memory}---information that is available to researchers today but was hidden from the public at the time.

 \begin{figure}[t]
  \centering
  \includegraphics[scale=.25]{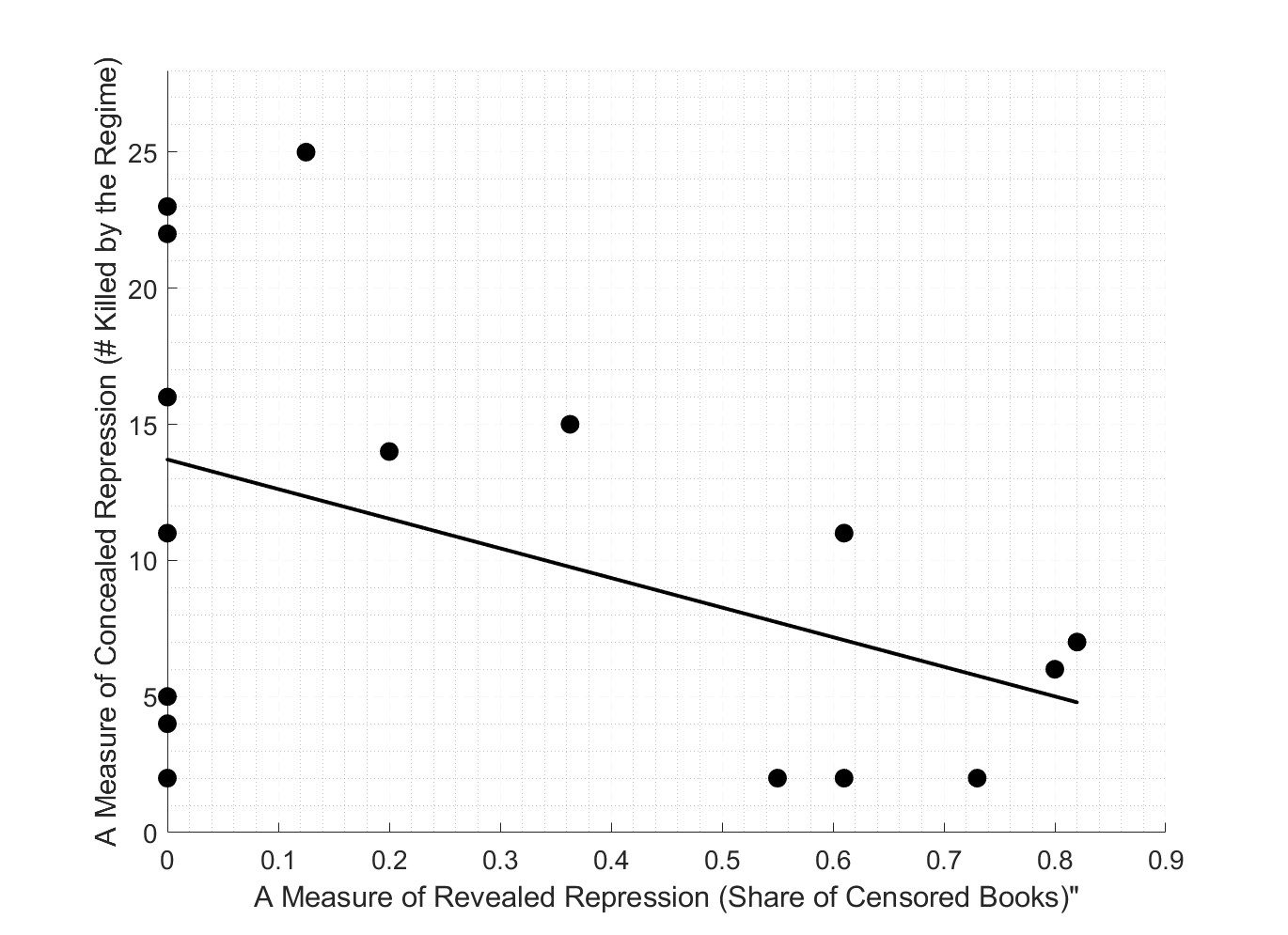} 
  \caption{Correlation of Measures of Concealed and Revealed Repression in the Brazilian Regime, 1964-1979. Black line presents the best fitting line.}
\label{fig1}
\end{figure}

The regime did not conceal all forms of repression. For example, it prevented the publication of articles and books that were not politically aligned with the regime, formalized in  Law~5250/67 and Decree-Law~1077/70 \autocite{memory}. The regime made little effort to conceal such repressive acts at the time, to the extent that newspaper pages would occasionally appear blank, or replaced by poems and cooking recipes---for examples, see \cite{oglobo} and \cite{memorialdemocracia}.
As a measure of revealed repression, we use data from \cite{reimao} on the annual share of books by Brazilian authors that were submitted for approval but censored.\footnote{See \textcite{esbergapsr} for a similar measure of censored films in Chile under Pinochet. The gradual re-opening (``abertura'') began in 1979, featuring the Amnesty Law of 1979, and the repeal of the Institutional Act 5, which had supported human rights violations  \autocite{skidmore,reimao}. Consequently, the extent and procedures of censorship changed after 1979, so that censorship data for books from \cite{reimao} for 1980-1985 are either missing or show no books submitted to censorship---see also \cite{ventura}.}

Figure~\ref{fig1} plots the relationship between these measures of concealed and revealed repression. The correlation coefficient is $-0.45$, and we reject the null hypothesis of zero correlation at the 10\% level. This negative association is  consistent with analyses in the popular press \autocite{opcao}.
The negative correlation between concealed and revealed repression is substantively consequential: The sign of the correlations or effects of the study may be reversed because total repression may move in the opposite direction of revealed repression.

\subsubsection{Measurement Problem Due to Deterrence}

Next, consider the second problem. Even if regimes did not conceal repression, measures of repression based on incidences of observed repression do not account for the deterrent effect of repression. \cref{cor:activist-selection} highlights the problem.

\begin{corollary}\label{cor:activist-selection}
        If the activist's costs of repression $r$ are almost always negligible, then the activist almost always organizes in any equilibrium. Formally, if $\Gamma(x)\rightarrow \mathbf{1}_{\{x>0\}}$, then $\gamma\rightarrow 1$ in any equilibrium.\footnote{That is, consider a sequence $\{ \Gamma_n \}$ of distributions satisfying Assumption 1.6 and let $\gamma_n$ be an equilibrium probability that the activist organizes when their cost-to-benefit distribution is $\Gamma_n$. If $\lim_{n \to \infty} \Gamma_n(x) =\mathbf{1}_{\{x>0\}}$, then $\lim_{n \to \infty} \gamma_n = 1.$}
    \end{corollary}

\noindent However, \cref{prop:equilibrium} shows that $\gamma<1$ in equilibrium. That is, the threat of incurring repression costs reduces the likelihood of the activist challenging the state by $1-\gamma>0$.

\cref{cor:activist-selection} also reflects an underlying assumption of our setting (embedded in $b>0$) that almost always there are groups with grievances, making demands on the government. This feature captures a finding of social movements literature that ``discontent is ever-present'' \autocite[p.~251]{jenkins1977insurgency} and ``grievances\dots are a fairly permanent and recurring feature of historical landscape'' \autocite[p.~298]{oberschall1978theories}, and will surface in form of demands on the state unless prevented by the political structure  \autocite{tilly1978mobilization, mcadam1999political, tarrow2011power}.

While the measurement problem due to concealment has been mostly overlooked,\footnote{\textcite{AndirinEtAl} emphasize that preventive repression ``is inherently less public, and less visible'' (p.~2), but do not analyze concealed repression more broadly.} the literature has identified the measurement problem due to deterrence \autocite{ritter-policy-2014,conrad_contentious_2019}, though it does not analyze its implications for the measurement of repression.

\section{Measuring Repression}\label{sec:measuring repression}

In this section, we introduce two indices of repression and leverage our theoretical framework to address the measurement challenges discussed in \cref{subsec:measurement problem}. We then present an empirical approach to estimate these indices. As a proof of concept, we implement this approach to construct two indices of total repression for Russia from 2020 to 2025 and compare them with existing measures of repression. We conclude by discussing the feasibility of obtaining reliable public opinion data in autocracies. 

\subsection{Indices of Repression}\label{subsec:indices or repression}

The  analysis  in \cref{subsec:measurement problem} highlights two problems:  $\Pr(\text{revealed repression})$ is a poor proxy for $\Pr(\text{total repression})$, and these probabilities do not capture that activists may not mobilize for fear of repression. The first stems from regime's strategic decision, the second from the activist's. 

We propose two indices of repression that aim to account for the strategic behavior of activists and the regime. Our first index focuses on the repression of all activists: 
\begin{eqnarray}\label{eqn:definition-of-R}
    R&:=&\prob(\text{activist organizes absent repression})-\prob(\text{activist organizes})\notag \\
    &\quad\quad +&\prob(\text{activist organizes})\cdot \prob(\text{total repression} \sep \text{activist organizes})\notag\\
    &=&1-\prob(\text{activist organizes}) +\prob(\text{total repression}).
    \end{eqnarray}
    The first line captures that activists may strategically refrain from organizing to avoid state repression, where we interpret ``absent repression'' as negligible costs of repression to the activists ($r\approx 0$). The second line captures the regime's strategic decision to conceal repression, featuring $\prob(\text{total repression})$ as opposed to $\prob(\text{revealed repression})$. The equality follows from \cref{cor:activist-selection}. 
    
    The index takes values  in $[0,1]$:
    \begin{eqnarray*}
    0\leq R  \leq 1-\prob(\text{activist organizes})\cdot (1-\prob(\text{total repression} \sep \text{activist organizes}))
    \leq 1
    \end{eqnarray*}
If repression completely deters activist from organizing, so $\prob(\text{activist organizes})=0$, or if the regime always represses all organized activists, so $\prob(\text{total repression} \sep \text{activist organizes})=1$, the index assumes its upper bound value $R=1$. Conversely, if repression does not deter activists from organizing and the regime never represses organized activists, the index assumes its lower bound value $R=0$.
    
Our second index builds on the discussion of normative perspectives and legitimate coercion at the end of \cref{sec:analysis}, focusing on the repression of activists whose success benefits the public---we have referred to them as ``good activists''.
Analogously to \eqref{eqn:definition-of-R}, we define   
    \begin{eqnarray}\label{eqn:definition-of-RG}
     R_G:=q-\prob(\text{activist organizes and is good}) +\prob(\text{total repression of good activist}).
\end{eqnarray}

The remainder of this subsection shows how we can leverage equilibrium relationships to express these indices in terms of observable variables.  We begin by expressing the probability of total repression (i.e., both revealed and concealed) as a function of the probability of revealed repression. Fix an equilibrium with $(\gamma,\kappa,\widetilde{c})$. Recall that $q'$ is the public's belief, upon observing repression, that the activists are good.
By Bayes Rule, the public's posterior belief about the activists is
 \begin{equation}\label{eqn:q,q',kappa}
     \frac{q'}{1-q'}=\frac{q}{1-q}\kappa,
 \end{equation} 
 where $\kappa$ is the equilibrium likelihood ratio of revealed repression of organized good activists versus bad activists.
Combining \eqref{eqn:q,q',kappa} with \eqref{eqn:revealed-repression}, we have 
\begin{equation}\label{eqn:prob-publicize-and-H}
    \prob( \text{revealed repression})=\gamma\frac{1-q}{1-q'}(1-H(\widetilde{c})).
\end{equation}
Combining \eqref{eqn:prob-publicize-and-H} with \eqref{eqn:concealed-repression}, we can express the probability of concealed repression, which is unobservable, in terms of the probability of revealed repression, which is observable:
\begin{equation}\label{eqn:concealed-as-revealed}
   \prob(\text{concealed repression})=\gamma H(\widetilde{c})=\gamma-\frac{1-q'}{1-q}\ \prob( \text{revealed repression}).
\end{equation}  
Finally, combining \eqref{eqn:concealed-as-revealed} with \eqref{eqn:total repression basic formula} allows us to express $ \prob(\text{total repression})$ in terms of $\gamma$, $q$, $q'$, and $ \prob(\text{revealed repression})$. The analogous formulas for repression of good activists can be similarly derived. \cref{prop2:empirical-total} summarizes the results.

\begin{proposition}\label{prop2:empirical-total}
    In any equilibrium, 
    \[ 
    \prob(\text{total repression}) 
    =\gamma-\frac{q-q'}{1-q}\ \prob(\text{revealed repression}),
    \]
    and
    \[ 
    \prob\left(\ \substack{\text{total repression of} \\ \text{good activists}}\ \right) 
    =\gamma\ q-\frac{q-q'}{1-q}\ \prob\left(\text{revealed repression}\right),
    \]
    where $q>q'$ by equation \eqref{eqn:q'<q}.
\end{proposition}

 The parameters \( q \) and \( q' \) are the \textit{public's opinion (beliefs)} about the activists, the opposition, or dissidents (all of whom we have collectively referred to as the activist). Specifically, they reflect the public’s belief that the activists’ success would constitute an improvement over the status quo. In the next section, we discuss how these concepts can be estimated from population surveys that measure the public's belief about activists absent revealed repression (for $q$) and after revealed repression (for $q'$).

\cref{prop2:empirical-total} focuses on the components of our indices driven by the regime’s strategic behavior. The indices also incorporate components arising from activists’ strategic behavior. The next result expresses the indices in terms of public beliefs and the probability of revealed repression.

\begin{proposition}\label{prop3:repression-index}
    In any equilibrium,
    \begin{equation}
        R=1-\frac{q-q'}{1-q}\ \prob(\text{revealed repression})\ \ \mbox{and}\ \ R_G=q-\frac{q-q'}{1-q}\ \prob(\text{revealed repression}),\notag
    \end{equation}
    where $q>q'$ by equation \eqref{eqn:q'<q}. 
\end{proposition}

\cref{prop3:repression-index} is 
 the main theoretical result of the paper,  which we will use in our empirical approach and its implementation to construct empirical indices for Russia from 2020 to 2025.
The proof follows from combining equations \eqref{eqn:definition-of-R} and \eqref{eqn:definition-of-RG} with Propositions \ref{prop:equilibrium} and \ref{prop2:empirical-total}. Importantly, while $q$ is the public's prior in a given period, both $q'$ and $\prob(\text{revealed repression})$ are equilibrium objects.
For example,  $\prob(\text{revealed repression})\approx 0$ if $H(c)\approx \mathbf{1}_{\{c\geq 0\}}$, so that the concealment costs are small,  and the regime's cost of conceding to good activists, $\alpha_G$, is large. Then $R\approx 1$ even though $\prob(\text{revealed repression})\approx 0$, because the regime hides almost all its repressive acts. 

In a setting with ``aligned regime'' where $\alpha_G=0$, from \cref{remark:aligned regime}, $q'=0$. This, combined with equation \eqref{eqn:aligned regime} and \cref{prop3:repression-index}, implies:
 $R^{\text{aligned regime}}=1-\gamma q$  and $R^\text{aligned regime}_G=q(1-\gamma)$.
That $R^\text{aligned regime}_G > 0$ reflects the deterrent effect of activists’ uncertainty about whether they may be viewed as harmful by the regime and repressed. If, in addition, we assume that there is no such uncertainty, then good activists always organize and the aligned regime always concedes, yielding $\underline{R}^\text{aligned regime}_G = 0$, where the underline indicates that this value is the lower bound of the index in this environment.

\subsection{Empirical Implementation}\label{sec:empirical-implementation}

This section explains how to use observational data to estimate the indices of repression introduced in \cref{subsec:indices or repression}. 

A researcher wants to estimate indices $R$ and $R_G$, described in equations \eqref{eqn:definition-of-R} and \eqref{eqn:definition-of-RG}, in a consecutive sequence of time periods (e.g., a calendar month) indexed by $t=1,\cdots,T$. In each period, an equilibrium is played according to \cref{prop:equilibrium}, so that the indices can be expressed as described in \cref{prop3:repression-index}. In each time period $t$, the researcher has access to two types of datasets: 
\begin{enumerate}
    \item Public opinion surveys on the public's views about activists. In particular, at the beginning of each period $t$, the survey asks a random sample of $N_t$ citizens, indexed by $i\in\{1,\cdots,N_t\}$, about their views on the opposition or activists. This includes questions such as: ``Suppose the current opposition/activist/dissident group were to gain political power. On a scale from 0 to 100, how likely is it that your  condition would improve as a result?'' or ``On a scale from 0 to 100, how likely is it that the country  would benefit as a result?''\footnote{The exact questions for measuring the public's beliefs about  activist or opposition groups depend on the environment. For example, the \cite{anes} surveys ask about the Black Lives Matter movement after 2016, and AmericasBarometer adjusts the names of political parties and opposition groups by country in a given year \autocite{americasbarometer}. A literature develops methods for eliciting more accurate responses on sensitive topics, including list experiments and other survey designs \autocite{rosenfeld_empirical_2016}.}
\item  Data on repressive events at a higher frequency than periods $t=1,...,T$. In particular, let $d=1,...,D_t$ denote the sub-periods within time period $t$. For example, $t$ may be a calendar month and $d$ a day in month $t$. The researcher observes an indicator $e_{d,t} \in \{0,1\}$ for whether a repressive event occurred in sub-period $d$ of period $t$. 
\end{enumerate}

Let $R_t$ and $R_{G,t}$ denote indices $R$ and $R_G$ in period $t$, respectively.  We propose a multi-step approach to estimate these indices, using \cref{prop3:repression-index}, with each step providing a consistent estimator for a different term in the indices.

\paragraph{Step 1: Measuring Beliefs} Using respondent $i$’s answers to survey questions at the beginning of period $t$, the researcher constructs the respondent’s prior belief about whether the success of an opposition group is beneficial and should be supported---we refer to such oppositions as ``good''. Denote this belief by $q_{i,t}$. Let $q_t$ denote the population average of this belief: $q_t = \mathbb{E}[q_{i,t}]$. We assume that $q_t$ captures the prior belief of a representative member of the public, whom we refer to as ``the public'' in the model. 

If $e_{d,t}=1$ for some $d\in\{1,...,D_t\}$,  let $q'_{i,t}$ be respondent $i$'s posterior belief, upon observing repression, that the success of the opposition group is beneficial (the opposition is good) at the end of period $t$. Then, $q'_{i,t}=q_{i,t+1}$. Analogously, let $q'_t = \mathbb{E}[q'_{i,t}]$ and assume that $q'_t$ captures the posterior belief of the representative member of the public after observing repression in period $t$. We estimate $q_t$ and $q'_t$ by averaging beliefs across the respondents: 
\begin{equation}\label{eqn:annoyed}
\hat{q}_t = \frac{1}{N_t}\sum_{i=1}^{N_t} q_{i,t}\ \  \text{and}\ \ \hat{q}'_t = \hat{q}_{t+1}= \frac{1}{N_{t+1}} \sum_{i=1}^{N_{t+1}} q_{i,t+1}.
\end{equation}
Consistency of $\hat{q}_t$ and $\hat{q}_t'$ follows under weak regularity conditions that allow the application of a Law of Large Numbers to the mean of a random sample  as $N_t \to \infty$ \autocite[Theorem 3.1]{wooldridge2010econometric}.

 If $e_{d,t}=0$ for all $d\in\{1,...,D_t\}$, we are unable to estimate  $q'_t$. As we will see in the following steps, the estimated probability of revealed repression will be $0$, so that using \cref{prop3:repression-index}, the value of $q'_t$ will be inconsequential for the repression indices.

\paragraph{Step 2: Measuring Revealed Repression} Using data $e_{d,t}$ on incidences of repression, estimate the probability of revealed repression in period $t$, $\prob_t(\text{revealed repression})$, as
\begin{equation}\label{eqn:Pr-revealed-repression-estimator}
\hat{\prob}_t(\text{revealed repression}) = \frac{1}{D_t}\sum_{d=1}^{D_t} e_{d,t}.
\end{equation}
 The consistency of this estimator follows under weak regularity conditions as $D_t \to \infty$, which allow for correlation across sub-periods \autocite[Ch.~7]{hamilton}.
If $e_{d,t}=0$ for all $d=1,\ldots,D_t$, so that no repressive events are observed in period $t$, then $\hat{\prob}_t(\text{revealed repression})=0$.

\paragraph{Step 3: Measuring Repression Indices} The researcher has obtained consistent estimators for $q_t, q_t'$, and $ \prob_t(\text{revealed repression})$ from the previous steps, denoted $\hat{q}_t, \hat{q}_t',$ and $ \hat{\prob}_t(\text{revealed repression})$, respectively. Plugging in these estimates into the formulas in \cref{prop3:repression-index}, we obtain the estimators of the indices:
\begin{equation}\label{eqn:hat-indices}
        \hat{R}_t=1-\frac{\hat{q}_t-\hat{q}_{t+1}}{1-\hat{q}_t}\ \hat{\prob}_t(\text{revealed repression})\ \ \mbox{,}\ \ \hat{R}_{G,t} = \hat{q}_t-\frac{\hat{q}_t-\hat{q}_{t+1}}{1-\hat{q}_t}\ \hat{\prob}_t(\text{revealed repression}),
    \end{equation}
    where we substituted  $\hat{q}'_t = \hat{q}_{t+1}$ from \eqref{eqn:annoyed}. 

\cref{prop4:cmt} shows that these estimators are consistent for the true indices as the sample size grows.

\begin{proposition}\label{prop4:cmt}
Suppose that $\hat{q}_t \to_p q_t$ and $\hat{\prob}_t(\text{revealed repression}) \to_p \prob_t(\text{revealed repression})$ for all $t=1,...,T$ as $N_t \to \infty, D_t \to \infty$. Then, $\hat{R}_t \to_p R_t$ and $\hat{R}_{G,t} \to_p R_{G,t}$ where $\hat{R}_t, \hat{R}_{G,t}$ are defined in \eqref{eqn:hat-indices}.
\end{proposition}

In Steps 1 and 2 above, we provide simple consistent estimators for $q_t$, $q'_t$, and $\prob_t(\text{revealed repression})$. \cref{prop4:cmt} shows that researchers may instead use alternative consistent estimators, including weighted averages or predicted beliefs conditional on respondent-level observed characteristics such as age.

\cref{prop3:repression-index} and the formulas for $R$ and $R_G$ are valid only if $q>q'$. However, due to  noise in the estimates, it may occur in finite samples that $\hat{q}_t<\hat{q}'_t$ even when $q_t>q'_t$. Researchers wishing to avoid this can implement shape restrictions by imposing the inequalities $R_t \leq 1$ and $q_t>q'_t$ in estimation \autocite{chetverikov}.

\subsection{Proof of Concept: Repression Indices for Russia, 2020--2025}\label{sec:Russia}

In this section, as a proof of concept, we apply our empirical approach to compute the repression indices $R_t$ and $R_{G,t}$ in Russia for the period 2020--2025 at both monthly and annual frequencies.

\paragraph{Data} For data on the public's beliefs about the opposition, we use public opinion surveys from the Russian Public Opinion Research Center \autocite{vciom}. VCIOM is a Russian public opinion firm that runs monthly public opinion surveys on political and social questions from a representative sample of the population.
\footnote{Scholars typically obtain public opinion data from Levada Center \autocite[e.g.][]{putin_endogenous_2024,rosenfeld2017reevaluating}, VCIOM \autocite{kirill_public_2018,kizilova_rally_2024}, the Foundation for Public Opinion \autocite{rosenfeld_popularity_2018}, and through their own surveys \autocite{Tertytchnaya-Lankina-2020,zakharov_effects_2024,frye2023,krishnarajan}. For purposes of demonstrating the implementation and feasibility of our approach, we use publicly available VCIOM data, available from \href{https://wciom.com/our-news/ratings/social-institutions}{VCIOM website} and the \href{https://web.archive.org/web/20240528004231/https://wciom.com/fileadmin/user_upload/ratings/social_institutions.xls}{internet archives} (retrieved February 28, 2026), while recognizing the connections between VCIOM and the Russian regime \autocite[e.g.][204-205]{kirill_public_2018}.}

\begin{figure}[t]
  \centering
  \includegraphics[width = \textwidth]{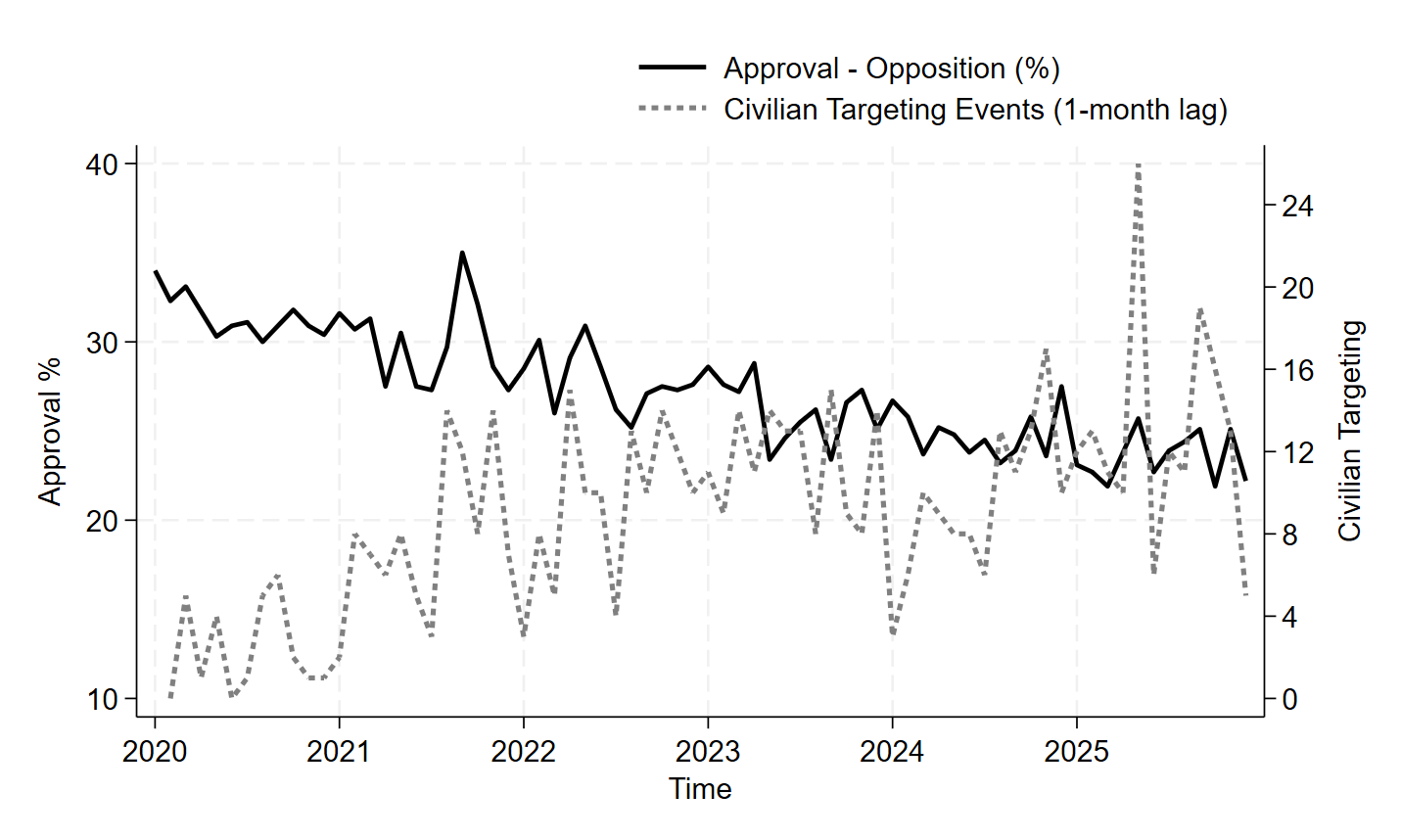} 
  \caption{Monthly opposition approval ratings (solid/black) and counts of revealed repression incidents (dashed/gray)}
\label{fig:approve&targeting}
\end{figure}

The surveys include a question, ``Do you generally approve or disapprove of the opposition?'' measured on a scale from 0 to 100, with higher numbers indicating higher approval.
The monthly survey responses averaged across the respondents from 46 Russian regions are publicly available from 2013 to 2025.
The approval rating for each month is released at the end of that month. We use that approval rating as our estimate of public opinion about the opposition at the end of that month. Because the data are already reported as approval ratings for the population, we take the approval rate for the month preceding the start of period $t$ as $\hat q_t$ in \eqref{eqn:annoyed}. For example, when $t$ refers to the first quarter of 2025, we take the approval rate of December 2024 as $\hat q_{t=2025Q1}$. 
Figure \ref{fig:approve&targeting} plots the monthly time series of public approval of the opposition from January 2020 to December 2025. The highest value is in September 2021 around the ``Fair Election Protests" \autocite{agence_france-presse_russias_2021}.

For repressive events, we use the variable \texttt{civilian targeting} from the Armed Conflict Location and Event Data (ACLED) \autocite{acled}.\footnote{ACLED is commonly used in the literature to measure protest and repression \autocite{zhukov,grasse,ives,rossdale,carter_propaganda_2021,anoll_protest_2025,brancati_locking_2023,christensen_concession_2019}.}
\texttt{Civilian targeting} registers repressive acts targeted at civilians at a daily frequency.
Repressive acts include attacks, sexual violence, abduction or forced disappearances, and excessive force against protesters. We focus on events where one of the actors is civilians in Russia (e.g.,  journalists, Navalny supporters, and communists) and the other the Russian state, e.g., Police Officers or Military.\footnote{Results are nearly identical if, instead, we use the variable \texttt{interaction 17: State forces-Civilians.}}
Because ACLED data are coded from news articles or other public communications \autocite{acled}, they reflect instances of repression revealed to the public. As of March, 2026, the daily data on repressive events in Russia are available from 2020 to 2025. Figure \ref{fig:approve&targeting} plots the monthly count of repression incidents.\footnote{\label{fn:qqprime}The correlation between the lag of monthly counts of repression incidents and the approval of the opposition is $-0.475$. This negative correlation is consistent with the model's prediction that revealed repression leads to the public updating negatively about the activists ($q'<q$) and with the literature on public opinion about dissidents following state repression \autocite{tertytchnaya_this_2023,pechenkina_how_2019}.} The large spike in April 2025 corresponds to increased repressive events in Chechnya after an attack on a police post \autocite{moscowtimes_2025,memorial_center}.

We construct the revealed repression indicator $e_{d,t}$ by setting $e_{d,t} = 1$ if \texttt{civilian targeting} is strictly positive on day $d$ in period $t$, and $e_{d,t} = 0$ otherwise.

\begin{figure*}[t!]
    \centering
    \begin{subfigure}[h]{0.5\textwidth}
            \includegraphics[width=\textwidth]{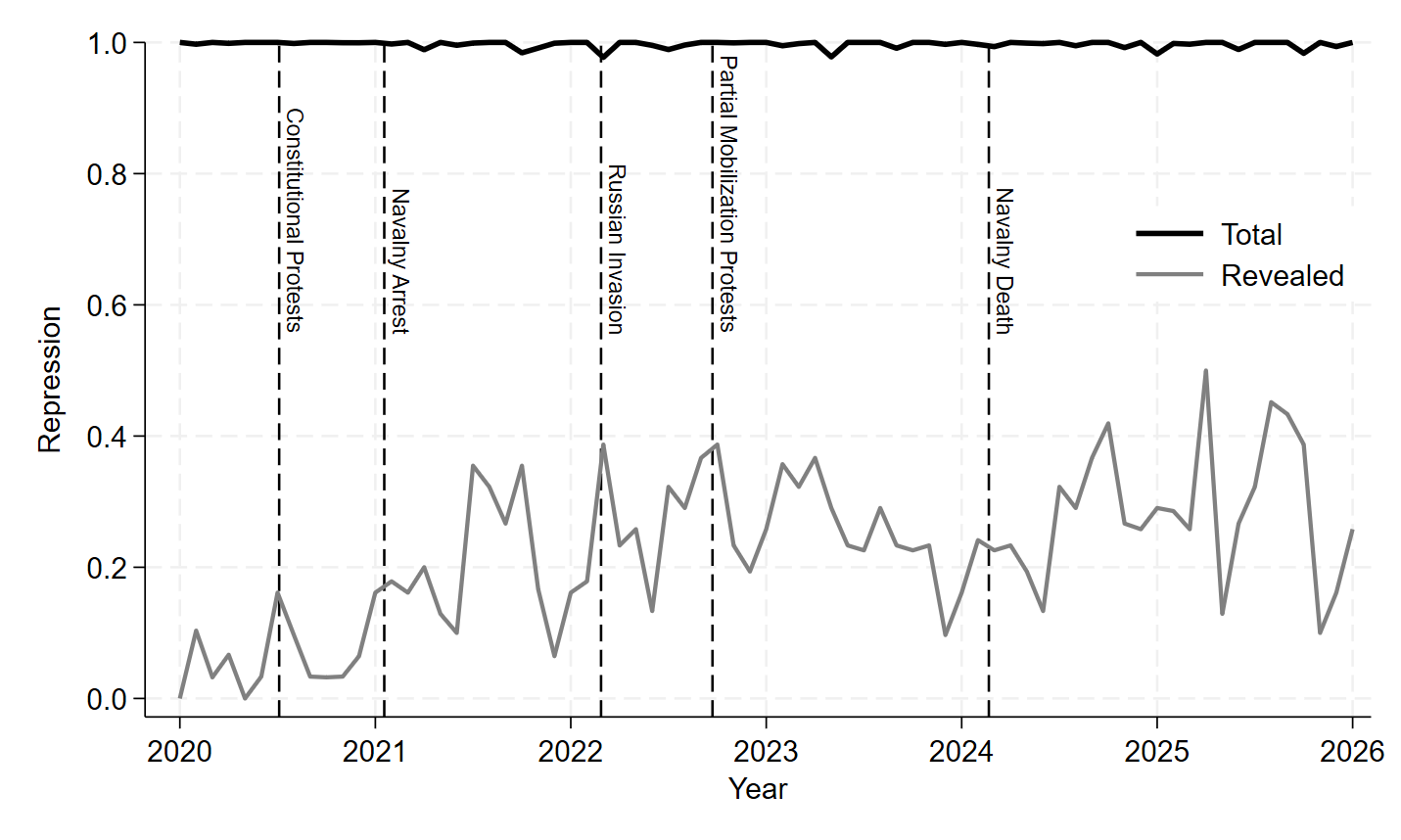}
    \end{subfigure}%
      ~
    \begin{subfigure}[h]{0.5\textwidth}
            \includegraphics[width=\textwidth]{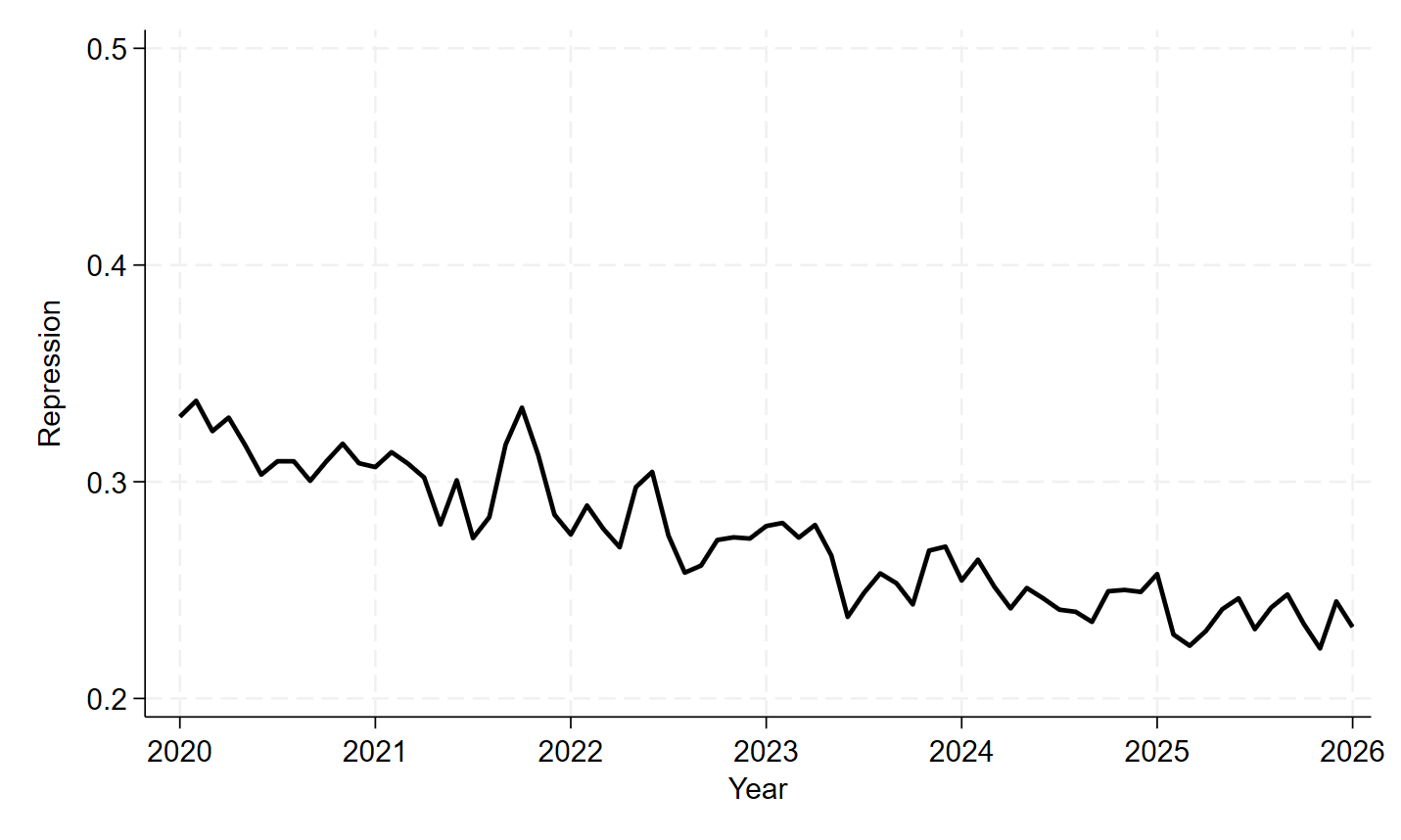}
    \end{subfigure}       
    \caption{Left panel: The black curve is the estimated index of repression $\hat R_t$; the gray curve is the estimated measure of revealed repression $\hat{\prob}_t(\text{revealed repression})$. Right panel: the estimated index of the repression $\hat R_{G,t}$; the corresponding measure of revealed repression requires estimating $\gamma$ for each period and is not provided.}
\label{totalrep}
\end{figure*}

\paragraph{Estimation} We begin by estimating the repression indices at the monthly frequency. Using equation \eqref{eqn:Pr-revealed-repression-estimator}, our estimate of the probability of revealed repression in period $t$, $\hat{\prob}_t(\text{revealed repression})$, is the number of days with incidents of revealed repression divided by the total number of days in period $t$, where each period corresponds to a calendar month. This measure is plotted in the left panel of Figure \ref{totalrep}. The estimated probability of revealed repression remains below 50\%, as most days do not feature observed repressive events. Moreover, as noted above, the public opinion data already come in the form of estimates $\hat{q}_t$.
Our estimates of the indices $\hat{R}_t$ and $\hat{R}_{G,t}$ are obtained by substituting $\hat{q}_t$ and $\hat{\prob}_t(\text{revealed repression})$ into \eqref{eqn:hat-indices} and imposing an upper bound of one.

\cref{totalrep} depicts the estimated indices. The estimated repression index $R$, measuring total repression, remains close to its maximum throughout the period, while the measure of revealed repression tends to increase. This increase in revealed repression is consistent with reports that the regime has increasingly restricted the freedom of assembly and expression in this period \autocite{noauthor_persecution_2025,amnesty24,kruope_disrupted_2025}. As expected, salient political events, such as the constitutional amendments \autocite{the_moscow_times_timeline_2020}, Navalny's arrest \autocite{roth_alexei_2021} and death \autocite{reevell_alexei_2024}, and the Russian invasion of Ukraine \autocite{noauthor_russia_2022}, correspond to relatively sharp increases in the measure of revealed repression.

In contrast, the estimated repression index $R_G$, which measures repression against activists whose success is beneficial from the public's perspective, declines over time. This index captures repression directed at activists whom the public would prefer not to be repressed. Combined with $R$ remaining near its maximum, this pattern suggests that the public perceives the regime as repressing ``good'' activists less and ``bad'' activists more. As repression in Russia has increasingly targeted anti-war opponents \autocite{noauthor_persecution_2025}, these trends may reflect how the public perceives the legitimacy of repressing different groups. We emphasize, however, that these estimates are based on VCIOM public opinion data and should be viewed as a proof of concept for our approach rather than a substantive analysis of the interactions between citizens and the state in Russia.

 \begin{figure}[t]
  \centering
  \includegraphics[width = \textwidth]{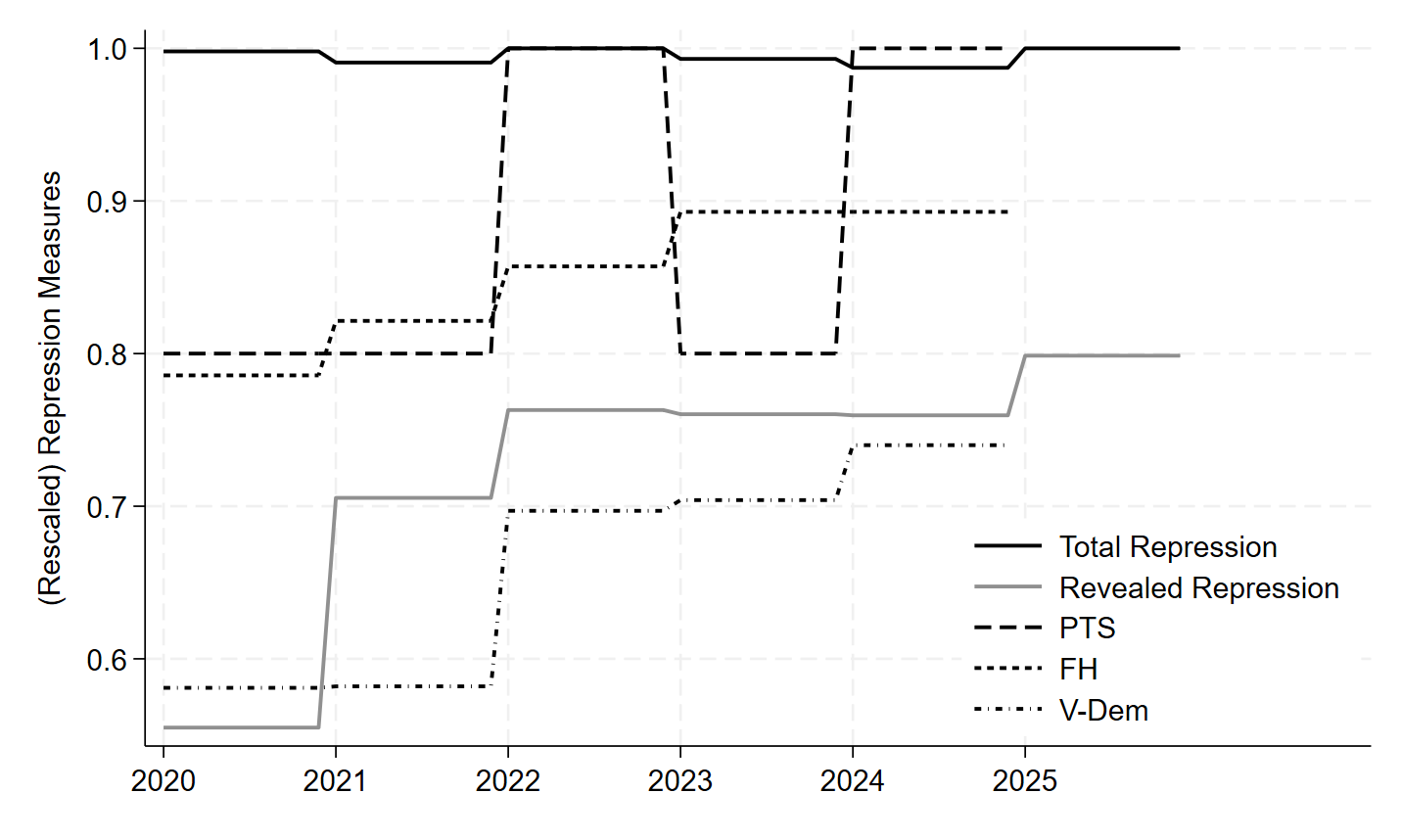} 
  \caption{Comparisons of the measures of total repression  and revealed repression (shifted) with the PTS, V-Dem, and FH indices.}
\label{fig:comparison}
\end{figure}

As discussed in \cref{sec:existing-measures}, scholars and human rights organizations have developed several indices for measuring repression. \cref{fig:comparison} depicts three indices for Russia over our time frame from the Political Terror Scale (PTS), Varieties of Democracy (V-Dem), and Freedom House (FH). 
Specifically, we use the PTS index based on Amnesty International reports, the V-Dem Civil Liberties Index, and the combined FH indices for Freedom of Expression \& Belief and Associational \& Organizational Rights.\footnote{The PTS index is a discrete measure with range $\{1,\cdots,5\}$. The V-Dem Civil Liberties Index (\texttt{v2x\_civlib}) is continuous with range $[0,1]$. The FH indices for Freedom of Expression \& Belief and Associational \& Organizational Rights are discrete measures with ranges $\{0,\cdots,16\}$ and $\{0,\cdots,12\}$, respectively. Because the FH indices are constructed additively, we divide the sum of these two indices by $28$. The most relevant sub-indices of the V-Dem Civil Liberties Index are the V-Dem Physical Integrity Rights Index (\texttt{v2x\_clphy}), which exhibits a pattern similar to PTS, and the V-Dem Political Civil Liberties Index (\texttt{v2x\_pcivlib}), which exhibits a pattern similar to the FH measure.}
We rescale these indices to the interval $[0,1]$ to facilitate comparison.
Scholars commonly use these and similar indices from the PTS \autocite[e.g.][]{davenport_stopping_2022,neuenkirch_impact_2016,fariss_respect_2014}, V-Dem \autocite[e.g.][]{cope_empirical_2019,tiscornia_police_2024,rubin_terrorism_2021}, and FH \autocite[e.g.][]{clay_human_2025,armstrongii_stability_2011,hendrix_when_2013}.
Because these indices are reported annually, measures whose values for 2025 had not been released as of March 2026 end in 2024. 
For comparability, we also present the estimated measures of total repression, the $R$ index, and revealed repression, $\prob(\text{revealed repression})$, at an annual frequency. To facilitate visual comparison of patterns, we shift the estimated measure of revealed repression upward by $0.5$.

The estimated measure of revealed repression exhibits an increasing pattern similar to the FH measure and, in particular, to the V-Dem index over this time frame. While revealed repression increases, the measure of total repression (the $R$ index) remains close to its maximum throughout.
This suggests that focusing on measures of revealed repression may underestimate the level of total repression and obscure its underlying patterns. For example, while in 2020 the Russian state may have engaged in less revealed repression than in 2024, the measure of total repression indicates that the overall level of repression was similar once both revealed and concealed repression are taken into account.

\subsection{Public Opinion Data in Autocracies}

This section assesses the feasibility of obtaining reliable public opinion data about opposition support in autocratic settings more broadly.

Scholars and research organizations routinely conduct surveys on politically sensitive topics in non-democratic regimes. Cross-national projects such as the \href{https://www.arabbarometer.org/}{Arab Barometer} field nationally representative surveys across the Middle East and North Africa, eliciting views on governments, civil society organizations, and freedoms of expression, assembly, press, and religion. Similar surveys include the \href{https://www.afrobarometer.org/}{Afrobarometer}, \href{https://www.vanderbilt.edu/lapop/about-americasbarometer.php}{AmericasBarometer}, and \href{https://www.worldvaluessurvey.org/}{World Values Survey}. In addition, country-specific initiatives collect data in authoritarian environments, including China (e.g., the Harvard Ash Center \autocite{ash}) and Iran, where respondents are asked about their preferences for alternative political systems and their support for protesters (e.g., \href{https://gamaan.org/}{GAMAAN} \autocite{gamaan,MalekiTamimiArab2026IranProtests}; see also \href{https://www.voxnations.com/}{VoxNations}).

A substantial literature also relies on country-specific surveys to measure public views on sensitive topics in authoritarian regimes, including 
China \autocite{HUANG_2015,huang_pathology_2018,pan,ZhuEtAl2025}, Egypt \autocite{truex_implicit_2019}, Hong Kong \autocite{Cantoni-et-al-2019}, Moldova \autocite{popeleches_censorship_2021}, Saudi Arabia \autocite{Bursztyn-et-al-2020}, Ukraine \autocite{pop-eleches_protest_2022}, Zimbabwe \autocite{lebas}, and Russia \autocite{Tertytchnaya-Lankina-2020,rosenfeld2017reevaluating,frye2023,krishnarajan}, measuring, e.g.,  public attitudes toward opposition figures and protest activity \autocite{tertytchnaya_this_2023}.

The repressive nature of these regimes raises concerns about preference falsification \autocite{kuran1997private}. However, a growing methodological literature develops tools to mitigate such concerns, including indirect questioning techniques and survey design innovations \autocite{blair2012statistical,frye2023,frye2024sensitivity,rosenfeld_empirical_2016}. Moreover, recent evidence suggests that the extent of preference falsification may be more limited than commonly assumed \autocite{frye2023,Shen_Truex_2021}, and in some contexts, conventional survey measures may outperform list experiments designed to reduce misreporting \autocite{frye2023}.

\section{Conclusion}

We develop a model of protest and repression in which the regime can conceal repression. We identify two measurement problems: one arising from concealment and one from deterrence. We construct repression indices that account for these challenges, propose an approach to estimate them using observable variables, and demonstrate the feasibility of this approach in a proof-of-concept application to Russia, 2020--2025.

Several directions for future research stand out. On the empirical front, our paper provides a proof of concept for implementing the proposed approach at a sub-annual frequency. A natural next step is to extend the analysis across longer time horizons and a broader set of countries.
On the theoretical front, it would be valuable to analyze the more sophisticated strategies that regimes may adopt to manipulate public opinion. For example, regimes may commit to repression strategies through institutional means---such as establishing partially independent judiciaries or shaping the composition of security forces. How do such institutional arrangements affect state--citizen interactions? Do they solely benefit the regime, or can they be Pareto-improving? In particular, if regimes are concerned that repression signals the presence of organized activists, they may repress members of the general public who pose no threat, thereby muddying the public's inference about activist presence when repression is observed. This suggests a novel rationale for indiscriminate repression. These directions are left for future research.

\clearpage

\printbibliography

\newpage
\appendix
\setcounter{page}{1}

\begin{center}
{\Huge Appendix}\\[1em]
\end{center}

\section{Proofs}\label{sec:appendix-proofs}

 \begin{proof}[Proof of \cref{lemma1:eqm-revolutions}]
    The public's expected payoff from protesting is $\mu(G)\beta_G + \mu(B) \beta_B - \rho$, which is positive if and only if $\rho \leq\widetilde{\rho}(\mu):=\mu(G)\beta_G + \mu(B) \beta_B$.
\end{proof}

\begin{lemma}\label{lemma:regime-reveals-under-D1}
    Under \cref{A1}, the regime represses-and-reveals with a strictly positive probability in any perfect Bayesian equilibrium that survives D1.
\end{lemma}
\begin{proof}[Proof of \cref{lemma:regime-reveals-under-D1}]
    Consider an equilibrium in which the regime represses-and-reveals with probability $0$. Then, the equilibrium payoff of type-$(\theta,c)$ regime is $\max\{ 1-F(\widetilde{\rho}(\mu_{NN})) - c, 1-\alpha_\theta \}$, where $1-\alpha_G > 1-\alpha_B$ by \cref{A1}.\ref{A1-alphaG<alphaB}. By the D1 criterion, the public's posterior $\mu_{RR}$ following unexpected revealed repression assigns positive probability to the type ($G$ or $B$) that benefits most from the deviation. In the equilibrium under consideration, the lowest payoff is $1-\alpha_B$ (since $\overline{c} > \alpha_B$ by \cref{A1}.\ref{A1-suppH}) and achieved by type-$(B,\overline{c})$ regime. Hence, $\mu_{RR} = (0,1,0)$ and the payoff from revealed repression is $1-F(\beta_B)$. Since $1-F(\beta_B) > 1-\alpha_B$ by Assumptions \ref{A1}.\ref{A1-alphaG<alphaB} and \ref{A1}.\ref{A1-betaB-GalphaG}, deviating to repress-and-reveal is strictly profitable for the type-$(B, \overline{c})$ regime. Consequently, an equilibrium in which the regime represses-and-reveals with probability $0$ does not survive the D1 refinement.
\end{proof}

\begin{lemma}\label{lemma:gamma-positive}
    Under \cref{A1}, there is no equilibrium in which the activist organizes with probability $0$ or $1$.
\end{lemma}
\begin{proof}[Proof of \cref{lemma:gamma-positive}]
    Let
    \begin{eqnarray}
        \prob(\text{repress-and-reveal}) &:=\int \sum_{\theta\in\{ G,B \}} \sigma(\text{repress-and-reveal} \sep \theta,c) \mu_0(\theta) dH(c),\label{eqn:prob-repres-and-reveal}
        \\
        \prob(\text{repress-and-conceal}) &:=\int \sum_{\theta\in\{ G,B \}} \sigma(\text{repress-and-conceal} \sep \theta,c) \mu_0(\theta) dH(c),\label{eqn:prob-repress-and-conceal}
        \\
        \prob(\text{concede}) &:=\int \sum_{\theta\in\{ G,B \}} \sigma(\text{concede} \sep \theta,c) \mu_0(\theta) dH(c)\label{eqn:prob-concede}
    \end{eqnarray}
be the total probabilities of the regime's respective actions, where $\mu_0(G) = q$ and $\mu_0(B) = 1-q$ is the prior probability that the activist is good or bad. Also, let
\begin{align*}
    \prob(\substack{\text{activist's proposal} \\ \text{is implemented}})
    := & \prob(\text{repress-and-reveal}) F( \widetilde{\rho}(\mu_{RR}) )\\
    + &\prob(\text{repress-and-conceal}) F( \widetilde{\rho}(\mu_{NN}) ) + \prob(\text{concede}), \\
    \prob(\substack{\text{activist is} \\ \text{repressed}}) := &\prob(\text{repress-and-reveal}) + \prob(\text{repress-and-conceal}).
\end{align*}
Then, an activist with a cost-to-benefit ratio $r/b$ organizes only if 
    \begin{equation}\label{eqn:r/b-organize}
        \frac{r}{b} \leq \frac{\prob(\substack{\text{activist's proposal} \\ \text{is implemented}})}{\prob(\substack{\text{activist is} \\ \text{repressed}})}.
    \end{equation}
Throughout the paper, we let $\gamma$ denote the probability that the activist organizes. From equation \eqref{eqn:r/b-organize}, in equilibrium, that probability equals:
\begin{eqnarray}
    \gamma = \Gamma (\Phi(\gamma)), \label{eqn:gamma=GammaPhigamma} \\ 
    \Phi(\gamma) := \prob(\substack{\text{activist's proposal} \\ \text{is implemented}}) \; \big/ \; \prob(\substack{\text{activist is} \\ \text{repressed}}) \label{eqn:Phi(gamma)-definition},
\end{eqnarray}
where the right-hand side ratio in \eqref{eqn:Phi(gamma)-definition} is evaluated in an equilibrium in which the activist organizes with probability $\gamma$.

We first argue that $\gamma < 1$ in any equilibrium. Indeed, if $\gamma = 1$, then $1 = \Gamma (\Phi(1))$ by \eqref{eqn:gamma=GammaPhigamma}, so 
$\Phi(1) \to \infty$ by \cref{A1}.\ref{A1-suppGamma}. However, $\prob(\text{repress-and-reveal}) > 0$ in any equilibrium by \cref{lemma:regime-reveals-under-D1}, which implies that $\Phi(\gamma)$ is finite in every equilibrium, a contradiction.

Next we argue that $\gamma > 0$ in any equilibrium. On the contrary, suppose that the activist organizes with probability zero. By \cref{A1}.\ref{A1-suppGamma}, $\Gamma(x) = 0$ if and only if $x = 0$. Consequently, in an equilibrium with no organized activists we must have $\Phi(0) = 0 \iff \prob\left(\substack{\text{activist's proposal} \\ \text{is implemented}}\right) = 0$. 

Observe that in an equilibrium with $\gamma = 0$, we have $\mu_{NN} = (0,0,1)$ and thus $F( \widetilde{\rho}(\mu_{NN}) ) = 0$. Also, $\prob(\text{repress-and-reveal}) >0$ in any equilibrium by \cref{lemma:regime-reveals-under-D1}. Thus, $\prob\left(\substack{\text{activist's proposal} \\ \text{is implemented}}\right) = 0$ if and only if
\begin{equation}\label{eqn:remark3-proof-gamma<1}
    F( \widetilde{\rho}(\mu_{RR}) ) = 0 \quad \text{ and }\quad \prob(\text{concede}) = 0.
\end{equation}
Now, if $\prob(\text{concede}) = 0$, then the type-$(\theta,c)$ regime represses-and-reveals (and gets $1-F( \widetilde{\rho}(\mu_{RR}) )$) or represses-and-conceals (and gets $1-F( \widetilde{\rho}(\mu_{NN}) )-c$). Either way, its payoff does not depend on the activist's type $\theta$ so the regime represses-and-reveals each type of activist with an equal probability. Then, $\mu_{RR} = (q,1-q,0)$ by the Bayes rule, and $F(\widetilde{\rho}(\mu_{RR})) = F(\beta^e) > 0$ (since $F(\beta^e) > \alpha_G$ by \cref{A1}.\ref{A1-betaE-GalphaG}), which contradicts \eqref{eqn:remark3-proof-gamma<1}. 
\end{proof}

\begin{proof}[Proof of \cref{prop:equilibrium}]

Consider an equilibrium in which the activist organizes with probability $\gamma$. By \cref{lemma:gamma-positive}, $\gamma \in (0,1)$. 

A type-$(\theta,c)$ regime strictly prefers to repress-and-conceal if 
    \begin{equation*}
        1 - F(\widetilde{\rho}(\mu_{NN})) - c > \max\{ 1 - F( \widetilde{\rho} (\mu_{RR}) ), 1-\alpha_\theta \},
    \end{equation*}    
    which is equivalent to
    \begin{equation}\label{eqn:c<c_theta}
        c < \widetilde{c}_\theta:=1 - F(\widetilde{\rho}(\mu_{NN})) - \max\{ 1 - F( \widetilde{\rho} (\mu_{RR}) ), 1-\alpha_\theta \}.
    \end{equation}
    The regime is indifferent if $c = \widetilde{c}_\theta$ (a measure-zero event), and strictly prefers not to repress-and-conceal if $c > \widetilde{c}_\theta$. Since $0<\alpha_G<\alpha_B$ (\cref{A1}.\ref{A1-alphaG<alphaB}), equation \eqref{eqn:c<c_theta} implies: 
    \begin{equation}\label{eqn:c_G-c_B}
        \widetilde{c}_G\leq \widetilde{c}_B\leq 1.
    \end{equation}
    We first show that there is no equilibrium in which $\widetilde{c}_G< \widetilde{c}_B$. By contradiction, suppose that $\widetilde{c}_G< \widetilde{c}_B$. Then, from \eqref{eqn:c<c_theta},  $1-\alpha_G>1-F(\widetilde\rho(\mu_{RR}))$, so repress-and-reveal is strictly dominated if $\theta=G$.  Hence, only the bad activists are publicly repressed (since the regime represses-and-reveals with a positive probability by \cref{lemma:regime-reveals-under-D1}), the public's posterior belief upon observing revealed repression is $\mu_{RR} = (0,1,0)$, and the regime's expected payoff from repress-and-reveal is $1-F(\beta_B)$. Next, note that if $\widetilde{c}_G< \widetilde{c}_B$, then from \eqref{eqn:c_G-c_B}, $\widetilde{c}_G<1$.  Observe that now a regime of type $(\theta,c)$ such that $\theta=G$ and $c\in(\widetilde{c}_G,1)$ has a profitable deviation. From  \eqref{eqn:c<c_theta}, in any equilibrium, this regime type either represses-and-reveals or concedes, so that its payoff is $\max\{1-\alpha_G,1 - F( \widetilde{\rho} (\mu_{RR}) )\}$, which is $1-\alpha_G$ if $\widetilde{c}_G< \widetilde{c}_B$. If this type deviates from the prescribed strategy and represses-and-reveals, then its payoff is $1-F(\beta_B)$. That deviation is profitable since $1-F(\beta_B)>1-\alpha_G$ by \cref{A1}.\ref{A1-betaB-GalphaG}, a contradiction. Therefore, in any equilibrium, $\widetilde{c}_G= \widetilde{c}_B=\widetilde{c}$, meaning that the regime represses-and-conceals either type of activist if $c < \widetilde{c}$.

    Next, since $\widetilde{c}_G= \widetilde{c}_B=\widetilde{c}$, from \eqref{eqn:c<c_theta} we have $1 - F( \widetilde{\rho} (\mu_{RR}) ) \geq 1-\alpha_G > 1-\alpha_B$. We now show that there is no equilibrium in which $1 - F( \widetilde{\rho} (\mu_{RR}) ) > 1-\alpha_G$. If that were the case, the regime never concedes; in particular, the regime represses-and-reveals either type of activist when $c > \widetilde{c}$. Then, by the Bayes rule, $\mu_{RR} = (q,1-q,0)$ and the expected payoff from repress-and-reveal is $1 - F(\beta^e)$. However, by \cref{A1}.\ref{A1-betaE-GalphaG}, we have $1-F(\beta^e) < 1-\alpha_G$, meaning that deviating to concession is strictly profitable for the regime of type $(G,c > \widetilde{c})$, a contradiction. Therefore, in any equilibrium, 
    \begin{equation}\label{eqn:prop1-indif-reveal-concedeG}
    1-F(\widetilde{\rho}(\mu_{RR}))= 1-\alpha_G > 1-\alpha_B, 
    \end{equation}
    the regime with repression cost $c > \widetilde{c}$ represses-and-reveals the bad activist, and is indifferent between repress-and-reveal and conceding to the good activist.

    Next we characterize the probability with which the regime represses-and-reveals the good activist. Let
    \[
        \kappa 
        = \frac{\int \sigma( \textit{repress-and-reveal} \sep G,c ) dH(c) }{ \int \sigma( \textit{repress-and-reveal} \sep B,c ) dH(c) }
        = \frac{\int\limits_{\widetilde{c}}^{\overline{c}} \sigma( \textit{repress-and-reveal} \sep G,c ) dH(c) }{ 1-H(\widetilde{c}) }
    \]
be the likelihood ratio of repress-and-reveal of good over bad activist, derived using the arguments we made in the previous paragraph. Then, 
$\mu_{RR} = \left( \frac{\kappa q}{\kappa q + 1-q}, \frac{1-q}{\kappa q + 1 - q}, 0 \right)$ 
and equation \eqref{eqn:prop1-indif-reveal-concedeG} yields
    \begin{align*}
        1 - F( \widetilde{\rho}(\mu_{RR}) )  = 1-\alpha_G
        &\iff F\left(\frac{\kappa q}{\kappa q + 1-q} \beta_G
        + \frac{1-q}{\kappa q + 1 - q} \beta_B \right)= \alpha_G.
    \end{align*}
Therefore, 
\begin{equation}\label{eqn:kappa}
    \kappa = \frac{1-q}{q} \frac{F^{-1}(\alpha_G) - \beta_B}{\beta_G - F^{-1}(\alpha_G)},
\end{equation}
where $\kappa > 0 \iff F^{-1}(\alpha_G) > \beta_B$ (by \cref{A1}.\ref{A1-betaB-GalphaG}) and $\kappa < 1 \iff \beta^e > F^{-1}(\alpha_G)$ (by \cref{A1}.\ref{A1-betaE-GalphaG}).

Next, we find $\mu_{NN}$ and $\widetilde{c}$. From \eqref{eqn:c<c_theta} and \eqref{eqn:prop1-indif-reveal-concedeG}, we have 
\begin{equation}\label{eqn:prop1-proof-c-tilde}
    \widetilde{c} = \alpha_G - F(\widetilde{\rho}(\mu_{NN})) \in [0,\alpha_G].
\end{equation}
Given the regime's strategy, the total probability that it chooses to repress-and-conceal a type-$\theta$ activist is 
    \[
    \int\sigma(\text{repress-and-conceal} \sep \theta,c) dH(c) = H(\widetilde{c}) < 1,
    \] 
    since $\widetilde{c} \leq \alpha_G < \alpha_B < \overline{c}$ by Assumptions \ref{A1}.\ref{A1-alphaG<alphaB} and \ref{A1}.\ref{A1-suppH}. Using Bayes' rule, we find that 
    \begin{equation}\label{eqn:prop1-proof-muNN}
    \mu_{NN} = \left( \frac{\gamma H(\widetilde{c}) q}{\gamma H(\widetilde{c}) + 1-\gamma}, \frac{\gamma H(\widetilde{c}) (1-q)}{\gamma H(\widetilde{c}) + 1-\gamma}, \frac{1-\gamma}{\gamma H(\widetilde{c}) + 1-\gamma} \right), 
    \end{equation} 
    or, equivalently, $\mu_{NN} = (\gamma' q, \gamma' (1-q), 1-\gamma')$, where $\gamma' := \frac{ \gamma H(\widetilde{c})}{\gamma H(\widetilde{c}) + 1-\gamma} < \gamma$.
    
    To find the concealment cost threshold $\widetilde{c}$, we plug the value for $\mu_{NN}$ obtained in \eqref{eqn:prop1-proof-muNN} into \eqref{eqn:prop1-proof-c-tilde} and rearrange terms to get:
\begin{eqnarray}
             F\left(\frac{\gamma H(\widetilde{c}) q}{\gamma H(\widetilde{c}) + 1-\gamma} \beta_G
    + \frac{\gamma H(\widetilde{c}) (1-q)}{\gamma H(\widetilde{c}) + 1-\gamma} \beta_B  \right)=\notag
    \\
    F\left(\frac{\gamma H(\widetilde{c})}{\gamma H(\widetilde{c})+ 1-\gamma} \beta^e\right)= \alpha_G-\widetilde{c}. \label{eqn:c-tilde}
    \end{eqnarray}
We show that equation \eqref{eqn:prop1-proof-c-tilde} has a unique solution $\widetilde{c} \in (0,\alpha_G)$ by checking the boundary conditions for $LHS(\widetilde{c}):= F\left(\frac{\gamma H(\widetilde{c})}{\gamma H(\widetilde{c})+ 1-\gamma} \beta^e\right)$ and $RHS(\widetilde{c}):= \alpha_G-\widetilde{c}$, showing that both these functions are continuous and monotone, and invoking the intermediate value theorem. Indeed, we have that $LHS$ is continuous and increasing in $\widetilde{c}$; also, $LHS(0) =0$ (because $H(0) = 0$ by \cref{A1}.\ref{A1-suppH} and $F(0) = 0$ by \cref{A1}.\ref{A1-suppF}) and $LHS(\alpha_G) = F\left(\frac{\gamma H(\alpha_G)}{\gamma H(\alpha_G)+ 1-\gamma} \beta^e\right) > 0$ (since $H(\alpha_G) > 0$ by \cref{A1}.\ref{A1-suppH} and $F(\rho) > 0$ for any $\rho > 0$ by \cref{A1}.\ref{A1-suppF}). Furthermore, $RHS$ is continuous and strictly decreasing; also, $RHS(0) = \alpha_G$ and $RHS(\alpha_G) = 0$. Therefore, \eqref{eqn:prop1-proof-c-tilde} has a unique solution $\widetilde{c} \in (0,\alpha_G)$.

The last step of the proof involves showing existence of an equilibrium. Recall from the proof of \cref{lemma:gamma-positive} that the activist organizes whenever $r/b$ is below a threshold. Also, the equilibrium probability of organizing $\gamma$ must satisfy \eqref{eqn:gamma=GammaPhigamma}, which states that $\gamma = \Gamma(\Phi(\gamma))$, where $\Phi(\gamma)$ is the ratio of the probability that the activist's demands are implemented to the probability that the activist is repressed, defined in \eqref{eqn:Phi(gamma)-definition}. Hence, it remains to show that equation \eqref{eqn:gamma=GammaPhigamma} admits a solution $\gamma \in (0,1)$.

In an equilibrium wherein the activist organizes with probability $\gamma \in (0,1)$, we have:
\begin{align*}
    \prob(\text{repress-and-reveal}) &= (\kappa q+1-q) (1-H(\widetilde{c}(\gamma))), \\
    \prob(\text{repress-and-conceal}) &= H(\widetilde{c}(\gamma)), \\
    \prob(\text{concede}) &= (1-\kappa) q (1-H(\widetilde{c}(\gamma))), \\
    F(\widetilde{\rho}(\mu_{RR})) &= \alpha_G, \\
    F(\widetilde{\rho}(\mu_{NN})) &= \alpha_G - \widetilde{c}(\gamma),
\end{align*}
where the first three equalities follow from the regime's strategy and equations \eqref{eqn:prob-repres-and-reveal}, \eqref{eqn:prob-repress-and-conceal} and \eqref{eqn:prob-concede}, respectively; the fourth equation follows from \eqref{eqn:prop1-indif-reveal-concedeG} and the fifth equation follows from \eqref{eqn:prop1-proof-c-tilde}.

Plugging these values into \eqref{eqn:Phi(gamma)-definition}, we get:
\begin{equation}\label{eqn:Phi-of-gamma}
    \Phi(\gamma) = \frac{(\kappa q+1-q) (1-H(\widetilde{c}(\gamma))) \alpha_G + H(\widetilde{c}(\gamma)) (\alpha_G - \widetilde{c}(\gamma)) + (1-\kappa) q (1-H(\widetilde{c}(\gamma)))}{(\kappa q+1-q) (1-H(\widetilde{c}(\gamma))) + H(\widetilde{c}(\gamma))}.
\end{equation}
Let
\begin{eqnarray}
    X := \kappa q + 1 - q \in (1-q,1), \label{eqn:X-definition-bounds} \\
    Y :=X\alpha_G+(1-\kappa)q= (\kappa q + 1 - q) \alpha_G + (1-\kappa)q \in (\alpha_G, (1-q)\alpha_G + q), \label{eqn:Y-definition-bounds}
\end{eqnarray}
where the bounds are obtained from observing that $X$ is strictly increasing in $\kappa$, $Y$ is strictly decreasing in $\kappa$ (since $\alpha_G < 1$), and $\kappa \in (0,1)$ from \eqref{eqn:kappa}. Plugging \eqref{eqn:X-definition-bounds} and \eqref{eqn:Y-definition-bounds} into \eqref{eqn:Phi-of-gamma}, we get:
\begin{equation}\label{eqn:Phi-simplified}
\Phi(\gamma)
=
\frac{ (1-H(\widetilde{c}(\gamma))) Y 
       + H(\widetilde{c}(\gamma)) (\alpha_G - \widetilde{c}(\gamma)) }
     {(1-H(\widetilde{c}(\gamma))) X 
       + H(\widetilde{c}(\gamma))}.
\end{equation}

We have:
\begin{equation}\label{eqn:sup-phi}
\Phi(\gamma)
\leq
\frac{ (1-H(\widetilde{c}(\gamma))) Y 
       + H(\widetilde{c}(\gamma)) \alpha_G }
     {(1-H(\widetilde{c}(\gamma))) X 
       + H(\widetilde{c}(\gamma))}
       \leq
\frac{Y}{X} < \alpha_G + \frac{q}{1-q}
=:\overline{\Phi}
<\infty,
\end{equation}
where the first inequality follows from $\widetilde{c}(\gamma) \geq 0$ (from \eqref{eqn:c-tilde}). The second inequality follows from observing that the function $\xi(H) := \frac{(1-H)Y + H \alpha_G}{(1-H)X + H}$ attains its maximum on $[0,1]$ when $H = 0$ because it is strictly decreasing in $H$: indeed, $\xi'(H) = \frac{\alpha_G X - Y}{((1-H)X + H)^2} = -\frac{(1-\kappa) q}{((1-H)X + H)^2} < 0$. The third inequality follows from from \eqref{eqn:X-definition-bounds} and \eqref{eqn:Y-definition-bounds}.

We also have:
\begin{eqnarray}
\Phi(\gamma)
>
\frac{ (1-H(\widetilde{c}(\gamma))) Y }
     {(1-H(\widetilde{c}(\gamma))) X 
       + H(\widetilde{c}(\gamma))}
       > \notag \\ 
       (1-H(\widetilde{c}(\gamma))) Y 
       >
(1-H(\alpha_G))Y
> (1-H(\alpha_G)) \alpha_G
=:\underline{\Phi}
>0,\label{eqn:inf-phi}
\end{eqnarray}
where the first and the third inequalities follow from $\widetilde{c} < \alpha_G$ (from \eqref{eqn:c-tilde}), the second inequality follows from $X < 1$ (by \eqref{eqn:X-definition-bounds}), the fourth inequality follows from $Y > \alpha_G$ (by \eqref{eqn:Y-definition-bounds}), and the fifth inequality follows from $H(\alpha_G)<1$ (by \cref{A1}.\ref{A1-suppH}). 

Next, since $\widetilde{c}(\gamma)$ is the unique solution to \eqref{eqn:c-tilde} and both sides of that equation are jointly continuous in $(\widetilde{c},\gamma)$, it follows that $\widetilde{c}(\gamma)$ is continuous.  Thus, $\Phi$ is a continuous function of $\gamma$. Since $\Gamma$ is continuous, the function $\gamma \mapsto \Gamma(\Phi(\gamma))$ is continuous. Moreover, from \eqref{eqn:inf-phi} and the fact that $\Gamma$ is increasing (as a CDF), we obtain $\Gamma(\Phi(\gamma)) \geq \Gamma(\underline{\Phi})$ for all $\gamma \in (0,1)$. Then,
\begin{equation}\label{eqn:Gamma-left}
\lim_{\gamma\to 0^+} \Gamma(\Phi(\gamma))
\ge \Gamma(\underline{\Phi}) > \Gamma(0)=0,
\end{equation}
which follows from \eqref{eqn:inf-phi} and the full support \cref{A1}.\ref{A1-suppGamma}. Using the same argument and equation \eqref{eqn:sup-phi}, we obtain:
\begin{equation}\label{eqn:Gamma-right}
   \lim_{\gamma\to 1^-} \Gamma(\Phi(\gamma))
\le \Gamma(\overline{\Phi}) < 1, 
\end{equation}
Therefore, by the Intermediate Value Theorem, there exists 
$\gamma \in (0,1)$ that solves $\gamma = \Gamma(\Phi(\gamma))$. That completes the proof of \cref{prop:equilibrium}.
\end{proof}

\begin{proof}[Proof of \cref{cor:activist-selection}]

From Part 1 of \cref{prop:equilibrium}, in any equilibrium, the activist organizes with probability $\gamma \in (0,1)$, which solves $\gamma = \Gamma(\Phi(\gamma))$. It follows that as $\Gamma\rightarrow1$, all solutions must approach $1$.
\end{proof}

 \begin{proof}[Proof of \cref{prop3:repression-index}]
We have:
\begin{eqnarray}
R &=&1-\gamma+\prob(\text{total repression}) \notag \\
  &=&1-\gamma + \gamma-\frac{q-q'}{1-q}\ \prob(\text{revealed repression})\notag\\
  &=&1-\frac{q-q'}{1-q}\ \prob(\text{revealed repression}).
\end{eqnarray}
where the first equality follows from substituting $\gamma = \prob(\text{activist organizes})$ into \eqref{eqn:definition-of-R} and the second equality follows from \cref{prop2:empirical-total}. Analogously, the index of repression of good activist becomes
\begin{eqnarray}
R_G &=&q-\gamma q  +\prob(\text{total repression of good activist})  \notag \\
  &=&q-\gamma q  + \gamma q -\frac{q-q'}{1-q}\ \prob(\text{revealed repression})\notag\\
  &=&q-\frac{q-q'}{1-q}\ \prob(\text{revealed repression}),
\end{eqnarray}
where the first equality follows from substituting $\gamma q  = \prob(\text{activist organizes and is good})$ into \eqref{eqn:definition-of-RG} and the second equality follows from \cref{prop2:empirical-total}.
\end{proof}

\begin{proof}[Proof of \cref{prop4:cmt}]
Fix an arbitrary $t=1,\ldots,T$. Define $z_t = \left(q_t, q_t', \prob_t(\text{revealed repression})\right)$ and $\hat{z}_t = \left(\hat{q}_t, \hat{q}_t', \hat{\prob}_t(\text{revealed repression})\right)$.

Because $\hat{q}_t \to_p q_t$, $\hat{q}_{t+1} = \hat{q}_t' \to_p q_t'$, and $\hat{\prob}_t(\text{revealed repression}) \to_p \prob_t(\text{revealed repression})$, we have that $\hat{z}_t \to_p z_t$  \autocite[p.~89]{hayashi}. Moreover, the function $g: [0,1]^3 \to [0,1]$, given by $g(z_t) = 1-\frac{q_t - q_t'}{1-q_t}\prob_t(\text{revealed repression})$, is continuous.
Thus, by the Continuous Mapping Theorem \autocite[Lemma 3.4, 3.6]{wooldridge2010econometric}, 
$\hat{R}_t = g(\hat{z}_t) \to_p g(z_t) = R_t$, provided that $R_t$ exists, which is true
under the model's assumption that $q_t \in (0,1)$ (see \cref{section:model}).
\end{proof}

\newpage

\section{Robustness}\label{sec:robustness}

\subsection{Exogenous Cost of Revealed Repression}

In the model of \cref{section:model}, revealing repression does not have direct costs. However, when repression is observed, the regime may incur costs beyond the public's near-term reaction. For example, the international community may sanction the regime \autocite{AndirinEtAl}, or the regime may incur long-term reputation costs \autocite{rasler1996concessions}. To capture these considerations, we augment the model by introducing a direct cost for revealed repression, so that engaging in repress-and-reveal now has a direct cost $K(\mu_{RR}(G))$ to the regime. We assume $K(\cdot)$ is weakly increasing with $K(0)=0$. Our main model then corresponds to the special case of $K=0$.

\begin{proposition}\label{prop:cost-of-repression-equilibrium}
Suppose that when the regime chooses to repress-and-reveal, it pays a  cost $K(\mu_{RR}(G))$, where $K(\cdot):[0,1]\rightarrow[0,\infty)$ is an increasing and continuous  function  with $K(0)=0$. Then, in every equilibrium, \cref{prop3:repression-index} holds.
\end{proposition}

\begin{proof}[Proof of \cref{prop:cost-of-repression-equilibrium}]
The equilibrium characterization analogous to \cref{prop:equilibrium} is described below in \cref{lemma:ext1-equilibrium}. Given \cref{lemma:ext1-equilibrium}, equations \eqref{eqn:concealed-repression}, \eqref{eqn:revealed-repression}, \eqref{eqn:total repression basic formula}, \eqref{eqn:q,q',kappa}, \eqref{eqn:prob-publicize-and-H}, \eqref{eqn:concealed-as-revealed} all hold. Consequently, Propositions \ref{prop2:empirical-total} and \ref{prop3:repression-index} also hold.
\end{proof}

\begin{lemma}\label{lemma:ext1-equilibrium}
    Suppose that when the regime chooses to repress-and-reveal, it pays a  cost $K(\mu_{RR}(G))$, where $K(\cdot):[0,1]\rightarrow[0,\infty)$ is an increasing and continuous  function  with $K(0)=0$. In every equilibrium: 
    \begin{enumerate}
      \item The activist organizes with probability $\gamma$, where $\gamma \in (0,1)$ solves $\gamma = \Gamma (\Phi(\gamma))$ and $\Phi(\gamma)$ is given by \eqref{eqn:Phi-of-gamma}.
    \item If $c < \widetilde{c}$,  the regime represses all organized activists and conceals that repression, where $\widetilde{c} \in (0,\alpha_G)$ is the unique solution to
    \begin{equation}\label{eqn:ext1-tilde-c}
            \frac{\gamma H(\widetilde{c})}{\gamma H(\widetilde{c}) + 1-\gamma}\ \beta^e  = F^{-1}(\alpha_G-\widetilde{c}).
    \end{equation}
    \item If $c > \widetilde{c}$, the regime publicly represses organized bad activists and represses organized good activists with probability less than 1. In particular, the  likelihood ratio of publicly repressing good versus bad organized activists 
    $\kappa \in (0,1)$ is the unique solution to
    \begin{equation}\label{eqn:ext1-kappa}
        F\left( 
        \beta_G \frac{\kappa q}{\kappa q + 1 - q} + \beta_B \frac{1-q}{\kappa q + 1 - q}
     \right)
     = \alpha_G
        - K\left( \frac{\kappa q}{\kappa q + 1 - q} \right).
    \end{equation}
\end{enumerate}    
\end{lemma}

\begin{proof}[Proof of \cref{lemma:ext1-equilibrium}]
First, observe that \cref{lemma:regime-reveals-under-D1} applies under the new specification of the regime's payoff under assumptions \ref{A1}.\ref{A1-alphaG<alphaB}, \ref{A1}.\ref{A1-betaB-GalphaG} and $K(0) = 0$. Specifically, the proof of \cref{lemma:regime-reveals-under-D1} changes only insofar that the payoff of type-$(B,\overline{c})$ regime from deviating to repress-and-reveal becomes $1-F(\beta_B)-K(0) = 1-F(\beta_B)$, while the rest of the proof applies as-is. \cref{lemma1:eqm-revolutions} and \cref{lemma:gamma-positive} hold because they do not rely on the regime's payoff. 

The proof of \cref{lemma:ext1-equilibrium} follows the same steps and uses the same arguments as the proof of \cref{prop:equilibrium}. Consider an equilibrium in which the activist organizes with probability $\gamma \in (0,1)$. The regime strictly prefers to repress-and-conceal if
\begin{equation}\label{eqn:ext1-c<c-tilde}
    c < \widetilde{c}_\theta:=1 - F(\widetilde{\rho}(\mu_{NN})) - \max\{ 1 - F( \widetilde{\rho} (\mu_{RR}) ) - K(\mu_{RR}), 1-\alpha_\theta \}.
\end{equation}
First, we can rule out equilibria in which $\widetilde{c}_G \neq \widetilde{c}_B$ and those in which the regime never concedes using the arguments from the proof of \cref{prop:equilibrium}. Thus, $\mu_{RR} = \left( \frac{\kappa q}{\kappa q + 1-q}, \frac{1-q}{\kappa q + 1 - q}, 0 \right)$ and from \eqref{eqn:ext1-c<c-tilde} we have:
\begin{eqnarray}
    1 - F(\widetilde{\rho}(\mu_{RR})) - K(\mu_{RR}) = 1-\alpha_G > 1-\alpha_B.\label{eqn:ext1-prekappa}
\end{eqnarray}
Then, equation \eqref{eqn:ext1-kappa} is obtained by rearranging terms in \eqref{eqn:ext1-prekappa} and substituting $\mu_{RR}$. To show that \eqref{eqn:ext1-kappa} has a unique solution $\kappa \in (0,1)$, we consider functions $LHS(\kappa) := F\left( 
        \beta_G \frac{\kappa q}{\kappa q + 1 - q} + \beta_B \frac{1-q}{\kappa q + 1 - q}
     \right)$ and $RHS(\kappa):=\alpha_G
        - K\left( \frac{\kappa q}{\kappa q + 1 - q} \right)$. 
We have that $LHS$ is continuous and strictly increasing in $\kappa$; also, $LHS(0) = F(\beta_B)$ and $LHS(1) = F(\beta^e)$. Furthermore, $RHS$ is continuous and decreasing in $\kappa$ (since $K$ is increasing); also, $RHS(0) = \alpha_G$ (since $K(0) = 0$) and $RHS(1) = \alpha_G - K(q) \leq \alpha_G$. The existence of the solution follows from the intermediate value theorem coupled with assumptions \ref{A1}.\ref{A1-betaE-GalphaG} and \ref{A1}.\ref{A1-betaB-GalphaG}, while uniqueness follows from monotonicity of left- and right-hand sides.

Next, $\mu_{NN}$ and $\widetilde{c}$ are the same as in the main model. Indeed, from \eqref{eqn:ext1-c<c-tilde} and the equality in \eqref{eqn:ext1-prekappa}, we obtain equation \eqref{eqn:prop1-proof-c-tilde} and thus equations \eqref{eqn:prop1-proof-muNN} and \eqref{eqn:c-tilde} survive as-is. 

The proof of equilibrium existence is exactly the same as in \cref{prop:equilibrium}.
\end{proof}

\subsection{Imperfect Concealment}

In the model of \cref{section:model}, when the regime chooses to repress-and-conceal, the regime's  concealment attempt succeeds, and the public does not observe repression. However, the regime may not always have access to such perfect means of concealment, and the news of repression may be revealed to the public despite the regime's concealment efforts. For example, citizens in Chile during the Pinochet dictatorship were generally kept in the dark about repressive activities due to the secretive nature of disappearances and persistent control over the media, but people who lived closer to military bases were more likely to observe or learn about repressive activities due to their more frequent use in those areas \autocite{bautista_geography_2023}.

To capture this feature, we augment the model by introducing a parameter $\chi$ that captures the effectiveness of the regime's concealment technology. In particular, when the regime chooses to conceal repression, it will be revealed to the public with probability $\chi\in [0,1)$. Our main model then corresponds to the special case of $\chi=0$.

\begin{proposition}\label{prop-ext2-main-result}
    Suppose that when the regime chooses to repress-and-conceal, the public observes repression with a probability 
\(
\chi \leq \dfrac{1 - H(\alpha_B)}{H(\alpha_B)}\dfrac{\kappa}{1-\kappa},
\)
where $\kappa$ is defined in \cref{prop:equilibrium}, and that $\alpha_B > F(\beta^e)$. Then, in every equilibrium, \cref{prop3:repression-index} holds.
\end{proposition}

\begin{proof}[Proof of \cref{prop-ext2-main-result}]
    From Lemmas \ref{lemma-ext2-equilibrium} and \ref{lemma-ext2-no-other}, under conditions of the proposition, every equilibrium is characterized by a triple $(\gamma,\kappa_\chi,\widetilde{c})$, where $\gamma,\kappa_\chi \in (0,1)$ and $\widetilde{c} \in (0, (1-\chi) \alpha_G )$. For brevity, denote $H:= H(\widetilde{c})$. From \eqref{eqn:ext2-muRR}, we have:
    \begin{eqnarray}
        \frac{q'}{1-q'} = \frac{\mu_{RR}(G)}{\mu_{RR}(B)} &= \frac{q}{1-q} \frac{\chi H + \kappa_\chi (1-H)}{\chi H + 1 - H} \implies \notag \\
        \chi H + \kappa_\chi(1-H) &= \frac{q'}{1-q'} \frac{1-q}{q} (\chi H + 1 - H). \label{eqn:ext2-plugin}
    \end{eqnarray}
    Furthermore, the ex-ante probability of revealed repression is:
    \begin{equation}
        \begin{split}\label{eqn:ext2-prob-revealed}
            \prob(\text{revealed repression}) &= \gamma[ q( \chi H + \kappa_\chi(1-H) ) + (1-q) (\chi H + 1 - H) ]  \\ 
            &= \gamma\frac{q'}{1-q'} (1-q)(\chi H + 1-H) + \gamma(1-q)(\chi H + 1-H)  \\ 
            &= \gamma ( \chi H + 1 - H ) \frac{1-q}{1-q'},
        \end{split}
    \end{equation}
    where the second equality plugs in the value for $\chi H + \kappa_\chi(1-H)$ from \eqref{eqn:ext2-plugin}. Rearranging terms in \eqref{eqn:ext2-prob-revealed}, we obtain:
    \begin{eqnarray}
        \frac{1-q'}{1-q} \prob(\text{revealed repression}) = \gamma ( \chi H + 1 - H ) \implies \notag \\
        \gamma - \frac{1-q'}{1-q} \prob(\text{revealed repression})
        = \gamma(1-\chi) H = \prob(\text{concealed repression}).\notag
    \end{eqnarray}
    
    Therefore, the ex-ante probability of total repression is
    \begin{align*}
        \prob(\text{total repression}) &= \prob(\text{revealed repression}) + \prob(\text{concealed repression}) \\
        &= \gamma - \frac{q-q'}{1-q} \prob(\text{revealed repression}),
    \end{align*} 
    and the ex-ante probability of total repression of good activists is
    \begin{align*}
        \prob\left(\ \substack{\text{total repression of} \\ \text{good activists}}\ \right)
        &= \prob(\text{total repression}) - \gamma (1-q) \\
    &=\gamma\ q-\frac{q-q'}{1-q}\ \prob\left(\text{revealed repression}\right),
    \end{align*}
    meaning that \cref{prop2:empirical-total} holds. Since the proof of \cref{prop3:repression-index} relies on \cref{prop2:empirical-total} and the definitions of $R$ and $R_G$ (given by \eqref{eqn:definition-of-R} and \eqref{eqn:definition-of-RG}, respectively), it also holds.
\end{proof}

\begin{lemma}\label{lemma-ext2-equilibrium}
Suppose that when the regime chooses to repress-and-conceal, the public observes repression with a probability $\chi \leq\dfrac{1 - H(\alpha_B)}{H(\alpha_B)}\dfrac{\kappa}{1-\kappa}$, where $\kappa$ is defined in \cref{prop:equilibrium}. There exists an equilibrium such that:
\begin{enumerate}
    \item The activist organizes with probability $\gamma$, where $\gamma \in (0,1)$ solves $\gamma = \Gamma (\Phi(\gamma))$ and $\Phi(\gamma)$ has the same functional form as \eqref{eqn:Phi-of-gamma} with $\kappa_\chi$ in place of $\kappa$.
    \item If $c < \widetilde{c}$,  the regime represses all organized activists and conceals that repression, where $\widetilde{c} \in (0,\alpha_G)$ is the unique solution to \eqref{eqn:ext2-c-tilde}.
    \item If $c > \widetilde{c}$, the regime publicly represses organized bad activists and represses organized good activists with probability less than 1. In particular, the  likelihood ratio of publicly repressing good versus bad organized activists 
    $\kappa_\chi \in (0,1)$ is the unique solution to \eqref{eqn:ext2-kappa}.
\end{enumerate}

    \end{lemma}

    \begin{proof}[Proof of \cref{lemma-ext2-equilibrium}]

    In the described equilibrium, the regime of type $(G,c > \widetilde{c})$ is indifferent between repress-and-reveal and concession (so that $1-F(\widetilde{\rho}(\mu_{RR})) = 1-\alpha_G$), the regime of type $(B, c > \widetilde{c})$ strictly prefers to repress-and-reveal (so that $1-F(\widetilde{\rho}(\mu_{RR})) = 1-\alpha_G > 1-\alpha_B$), the regime of type $(\theta,c < \widetilde{c})$ strictly prefers to repress-and-conceal. Finally, the regime of type $(G,\widetilde{c})$ is indifferent between all three actions:
    \begin{equation*}
        1-F(\widetilde{\rho}(\mu_{RR})) = 1-\alpha_G = 1-\widetilde{c} - \chi F(\widetilde{\rho}(\mu_{RR})) - (1-\chi) F(\widetilde{\rho}(\mu_{NN})),
    \end{equation*}
    so that:
    \begin{eqnarray}
        F(\widetilde{\rho}(\mu_{NN})) = \alpha_G - \frac{\widetilde{c}}{1-\chi}, \label{eqn:ext2-eqn1} \\ 
        F(\widetilde{\rho}(\mu_{RR})) = \alpha_G. \label{eqn:ext2-eqn2}
    \end{eqnarray}

    Given the regime's strategy, the public's posterior belief after no news is:
    \begin{equation}\label{eqn:ext2-muNN}
        \mu_{NN} = \left( \frac{\gamma q (1-\chi) H(\widetilde{c})}{D_{NN}}, \frac{\gamma (1-q) (1-\chi) H(\widetilde{c})}{D_{NN}}, \frac{1-\gamma}{D_{NN}} \right),
    \end{equation}
    where $D_{NN} = \gamma (1-\chi) H(\widetilde{c}) + 1 - \gamma$. Consequently, 
    \(
        \widetilde{\rho}(\mu_{NN}) = \frac{\gamma (1-\chi) H(\widetilde{c}) \beta^e}{\gamma (1-\chi) H(\widetilde{c}) + 1 - \gamma}
    \)
    and equation \eqref{eqn:ext2-eqn1} becomes:
    \begin{equation}\label{eqn:ext2-c-tilde}
        F \left( \frac{\gamma (1-\chi) H(\widetilde{c})}{\gamma (1-\chi) H(\widetilde{c}) + 1 - \gamma} \beta^e \right) = \alpha_G - \frac{\widetilde{c}}{1-\chi}.
    \end{equation}
    Now, let $LHS(\widetilde{c}) := F \left( \frac{\gamma (1-\chi) H(\widetilde{c})}{\gamma (1-\chi) H(\widetilde{c}) + 1 - \gamma} \beta^e \right)$ and $RHS(\widetilde{c}) := \alpha_G - \frac{\widetilde{c}}{1-\chi}$. Note that $LHS(\widetilde{c})$ is continuous and strictly increasing; also, $LHS(0) = 0$ and $LHS( (1-\chi)\alpha_G ) > 0$ by assumptions \ref{A1}.\ref{A1-suppH} and \ref{A1}.\ref{A1-suppF}. Furthermore, $RHS(\widetilde{c})$ is strictly decreasing; also, $RHS(0) = \alpha_G > 0$ and $RHS( (1-\chi)\alpha_G ) = 0$. Therefore, there exists a unique $\widetilde{c} \in (0, (1-\chi)\alpha_G)$ that solves equation \eqref{eqn:ext2-c-tilde}.

    Next, we work with equation \eqref{eqn:ext2-eqn2}. Given the regime's strategy, the public's posterior after revealed repression is:
    \begin{equation}\label{eqn:ext2-muRR}
        \mu_{RR} = \left( \frac{q( \chi H(\widetilde{c}) + \kappa_\chi (1-H(\widetilde{c})) )}{D_{RR}},
        \frac{(1-q)(\chi H(\widetilde{c}) + 1 - H(\widetilde{c}))}{D_{RR}},0
        \right),
    \end{equation}
    where $D_{RR} = q( \chi H(\widetilde{c}) + \kappa_\chi (1-H(\widetilde{c})) ) + (1-q)(\chi H(\widetilde{c}) + 1 - H(\widetilde{c}))$. Consequently, equation \eqref{eqn:ext2-eqn2} yields:
    \begin{eqnarray}
        F \left( \frac{q(\chi H(\widetilde{c}) + \kappa_\chi (1 - H(\widetilde{c})))\beta_G + (1-q)(\chi H(\widetilde{c}) + 1 - H(\widetilde{c})) \beta_B}{D_{RR}} \right) = \alpha_G \iff \notag \\
        \frac{\chi H(\widetilde{c}) + \kappa_\chi (1-H(\widetilde{c}))}{\chi H(\widetilde{c}) + 1 - H(\widetilde{c})} = \frac{1-q}{q} \frac{F^{-1}(\alpha_G) - \beta_B}{\beta_G - F^{-1} (\alpha_G)} = \kappa \in (0,1),\label{eqn:ext2-kappa}
    \end{eqnarray}
    where $\kappa \in (0,1)$ as shown in \eqref{eqn:kappa}. The left-hand side of \eqref{eqn:ext2-kappa} is strictly increasing in $\kappa_\chi$, ranging from $\frac{\chi H(\widetilde{c})}{\chi H(\widetilde{c}) + 1 - H(\widetilde{c})}$ at $\kappa_\chi = 0$ to $1$ at $\kappa_\chi = 1$. A unique solution $\kappa_\chi \in (0,1)$ thus exists if and only if $\frac{\chi H(\widetilde{c})}{\chi H(\widetilde{c}) + 1 - H(\widetilde{c})} < \kappa \iff \chi < \frac{1-H(\widetilde{c})}{H(\widetilde{c})}\cdot\frac{\kappa}{1-\kappa}$. Since $\widetilde{c} < \alpha_G < \alpha_B$, we have $H(\widetilde{c}) < H(\alpha_G) < H(\alpha_B)$, and $\chi \leq\frac{1 - H(\alpha_B)}{H(\alpha_B)}\frac{\kappa}{1-\kappa}$ therefore implies that $\chi < \frac{1 - H(\widetilde{c})}{H(\widetilde{c})}\frac{\kappa}{1-\kappa}$, so the solution to \eqref{eqn:ext2-kappa} exists and is unique.

    The last step of the proof of \cref{lemma-ext2-equilibrium} is showing existence of the fixed point for equation $\gamma = \Gamma (\Phi(\gamma))$. With imperfect concealment, when the regime chooses to repress-and-conceal, the public observes repression with probability $\chi$ and no news with probability $1-\chi$. We have:
    \begin{align*}
    \prob(\substack{\text{activist's proposal} \\ \text{is implemented}})
    ={}& \prob(\text{repress-and-reveal})\, F( \widetilde{\rho}(\mu_{RR}) )\\
    &+\prob(\text{repress-and-conceal})\,\big[\chi\, F( \widetilde{\rho}(\mu_{RR}) )
      + (1-\chi)\, F( \widetilde{\rho}(\mu_{NN}) )\big]\\
    &+ \prob(\text{concede}),\\[4pt]
    \prob(\substack{\text{activist is} \\ \text{repressed}})
    ={}& \prob(\text{repress-and-reveal}) + \prob(\text{repress-and-conceal}),\\[4pt]
    \prob(\text{repress-and-reveal}) &= (\kappa_\chi q+1-q) (1-H(\widetilde{c}(\gamma))), \\
    \prob(\text{repress-and-conceal}) &= H(\widetilde{c}(\gamma)), \\
    \prob(\text{concede}) &= (1-\kappa_\chi) q (1-H(\widetilde{c}(\gamma))),
    \end{align*}
    Plugging these values and equations \eqref{eqn:ext2-eqn1} and \eqref{eqn:ext2-eqn2} into \eqref{eqn:Phi(gamma)-definition}, we find that $\Phi(\gamma)$ has the same functional form as \eqref{eqn:Phi-of-gamma} with $\kappa_\chi$ in place of $\kappa$. Since $\kappa_\chi \in (0,1)$, the quantities $X_\chi := \kappa_\chi q + 1-q$ and $Y_\chi := X_\chi \alpha_G + (1-\kappa_\chi) q$ satisfy the same bounds as \eqref{eqn:X-definition-bounds} and \eqref{eqn:Y-definition-bounds}: $X_\chi \in (1-q,1)$ and $Y_\chi \in (\alpha_G, (1-q)\alpha_G + q)$. The rest of the proof that $\gamma = \Gamma(\Phi(\gamma))$ then follows the same steps as the corresponding part of the proof of \cref{prop:equilibrium}, with $\kappa_\chi$ replacing $\kappa$ throughout. The proof of \cref{lemma-ext2-equilibrium} is now complete.
\end{proof}

\begin{lemma}\label{lemma-ext2-no-other}
Suppose that when the regime chooses to repress-and-conceal, the public observes repression with a probability 
\(
\chi \leq \dfrac{1 - H(\alpha_B)}{H(\alpha_B)}\dfrac{\kappa}{1-\kappa},
\)
where $\kappa$ is defined in \cref{prop:equilibrium}, and that $\alpha_B > F(\beta^e)$. Then there exists no equilibrium other than the one characterized in \cref{lemma-ext2-equilibrium}.
\end{lemma}

\begin{proof}
Suppose, toward a contradiction, that there exists an equilibrium other than the one in \cref{lemma-ext2-equilibrium}. We consider three cases based on $F(\widetilde{\rho}(\mu_{RR}))$.

\medskip

\noindent\textit{Case 1: $F(\widetilde{\rho}(\mu_{RR}))<\alpha_G$.} Then $1-F(\widetilde{\rho}(\mu_{RR}))>1-\alpha_G>1-\alpha_B$, so repress-and-reveal yields a strictly higher payoff than concession for both types. Therefore, the regime represses-and-reveals both types at equal rates when $c>\widetilde{c}$, giving $\mu_{RR}(G)/\mu_{RR}(B)=q/(1-q)$ by Bayes' rule. This implies $F(\widetilde{\rho}(\mu_{RR}))=F(\beta^e)>\alpha_G$ by \cref{A1}.\ref{A1-betaE-GalphaG}, a contradiction.

\medskip

\noindent\textit{Case 2: $F(\widetilde{\rho}(\mu_{RR}))\in(\alpha_G,\alpha_B)$.} Since $1-\alpha_G>1-F(\widetilde{\rho}(\mu_{RR}))$, concession strictly dominates repress-and-reveal for type~$G$: the regime concedes to good activists whenever $c>\widetilde{c}_G$ and never represses-and-reveals them. Since $1-F(\widetilde{\rho}(\mu_{RR}))>1-\alpha_B$, repress-and-reveal strictly dominates concession for type~$B$: the regime represses-and-reveals bad activists whenever $c>\widetilde{c}_B$. Because the concealment payoff at $c=0$ does not depend on $\theta$ while good activists have a strictly better outside option ($1-\alpha_G>1-F(\widetilde{\rho}(\mu_{RR}))$), we have $\widetilde{c}_G<\widetilde{c}_B$. Moreover, $\widetilde{c}_B=(1-\chi)(F(\widetilde{\rho}(\mu_{RR}))-F(\widetilde{\rho}(\mu_{NN})))\leq (1-\chi)F(\widetilde{\rho}(\mu_{RR}))<F(\widetilde{\rho}(\mu_{RR}))<\alpha_B$.

The public observes repression of good activists only when concealment leaks (probability $q\chi H(\widetilde{c}_G)$), while it observes repression of bad activists through both leaked concealment and repress-and-reveal (probability $(1-q)[1-(1-\chi)H(\widetilde{c}_B)]$). Since $\widetilde{c}_G<\alpha_B$ and $\widetilde{c}_B<\alpha_B$, we have $H(\widetilde{c}_G)\leq H(\alpha_B)$ and $H(\widetilde{c}_B)\leq H(\alpha_B)$. Therefore,
\[
\frac{\mu_{RR}(G)}{\mu_{RR}(B)}
=
\frac{q\chi H(\widetilde{c}_G)}{(1-q)\bigl[1-(1-\chi)H(\widetilde{c}_B)\bigr]}
\leq
\frac{q\chi H(\alpha_B)}{(1-q)\bigl[1-(1-\chi)H(\widetilde{c}_B)\bigr]}
\]
\[
\leq
\frac{q\chi H(\alpha_B)}{(1-q)\bigl[1-(1-\chi)H(\alpha_B)\bigr]},
\]
where the first inequality replaces $H(\widetilde{c}_G)$ with the larger $H(\alpha_B)$ in the numerator, and the second replaces $H(\widetilde{c}_B)$ with $H(\alpha_B)$ in the denominator, which decreases it since $(1-\chi)H(\widetilde{c}_B)\leq (1-\chi)H(\alpha_B)$. By assumption,
$\chi \leq \frac{1-H(\alpha_B)}{H(\alpha_B)}\frac{\kappa}{1-\kappa}$,
which is equivalent to $\frac{\chi H(\alpha_B)}{1-(1-\chi)H(\alpha_B)}\leq \kappa$. Hence $\frac{\mu_{RR}(G)}{\mu_{RR}(B)}\leq \frac{q}{1-q}\kappa$.

Writing $r:=\mu_{RR}(G)/\mu_{RR}(B)$, we have $\widetilde{\rho}(\mu_{RR})=\frac{r\beta_G+\beta_B}{r+1}$, which is strictly increasing in $r$ since $\beta_G>\beta_B$. When
$r=\frac{q}{1-q}\kappa = \frac{F^{-1}(\alpha_G)-\beta_B}{\beta_G-F^{-1}(\alpha_G)}$
(substituting the definition of $\kappa$ from \eqref{eqn:kappa}), direct computation gives $\widetilde{\rho}(\mu_{RR})=F^{-1}(\alpha_G)$, so $F(\widetilde{\rho}(\mu_{RR}))=\alpha_G$. Since $r\leq \frac{q}{1-q}\kappa$ and $\widetilde{\rho}$ is increasing in $r$, we conclude $F(\widetilde{\rho}(\mu_{RR}))\leq\alpha_G$, a contradiction.

\medskip

\noindent\textit{Case 3: $F(\widetilde{\rho}(\mu_{RR}))\geq\alpha_B$.} Since $F(\widetilde{\rho}(\mu_{RR}))\geq\alpha_B>\alpha_G$, concession strictly dominates repress-and-reveal for type~$G$ (as in Case~2). Moreover, $F(\widetilde{\rho}(\mu_{RR}))\geq\alpha_B$ implies $1-F(\widetilde{\rho}(\mu_{RR}))\leq 1-\alpha_B$, so concession weakly dominates repress-and-reveal for type~$B$ as well. The concealment thresholds satisfy $\widetilde{c}_B-\widetilde{c}_G=\alpha_B-\alpha_G>0$ (since both types' outside option is concession and the concealment payoff is type-independent), so $\widetilde{c}_G<\widetilde{c}_B$.

Good activists are never repressed-and-revealed (as in Case~2), so the public observes repression of good activists only when concealment leaks (probability $q\chi H(\widetilde{c}_G)$). The public observes repression of bad activists through leaked concealment (probability $(1-q)\chi H(\widetilde{c}_B)$) and possibly through repress-and-reveal; the former is a lower bound, as repress-and-reveal can only increase the probability that the public observes repression of bad activists. Therefore,
\[
\mu_{RR}(G)
\leq
\frac{q\chi H(\widetilde{c}_G)}{q\chi H(\widetilde{c}_G)+(1-q)\chi H(\widetilde{c}_B)}
=
\frac{q H(\widetilde{c}_G)}{q H(\widetilde{c}_G)+(1-q) H(\widetilde{c}_B)}
\leq q,
\]
where the last inequality follows from $H(\widetilde{c}_G)\leq H(\widetilde{c}_B)$ (since $\widetilde{c}_G<\widetilde{c}_B$ and $H$ is increasing). Consequently, $\widetilde{\rho}(\mu_{RR})=\mu_{RR}(G)\beta_G+(1-\mu_{RR}(G))\beta_B\leq q\beta_G+(1-q)\beta_B=\beta^e$, so
\[
F(\widetilde{\rho}(\mu_{RR}))\leq F(\beta^e)<\alpha_B,
\]
where the last inequality holds by assumption. This contradicts $F(\widetilde{\rho}(\mu_{RR}))\geq\alpha_B$.

\medskip

Since all three cases yield contradictions, no equilibrium other than the one in \cref{lemma-ext2-equilibrium} exists.
\end{proof}

\end{document}